\documentclass[journal=jacsat,manuscript=article]{achemso}

\usepackage[version=3]{mhchem} % Formula subscripts using \ce{}
\usepackage[T1]{fontenc}       % Use modern font encodings
\usepackage{graphicx}% Include figure files
\usepackage{bm}% bold math
\usepackage{color}
\usepackage{amsmath, amsthm, amssymb, amsfonts}

\author{Shumpei Masuda}
\email{shumpei.masuda@aalto.fi}
\author{Kuan Y. Tan}
\author{Matti Partanen}
\author{Russell E. Lake}
\author{Joonas Govenius}
\author{Matti Silveri}
\altaffiliation{Research Unit of Theoretical Physics, University of Oulu, FI-90014 Oulu, Finland}

\affiliation{QCD Labs, COMP Centre of Excellence, 
Department of Applied Physics, Aalto University, 
PO Box 13500, FI-00076 Aalto, Finland}
\author{Hermann Grabert}
\affiliation{Department of Physics, University of Freiburg, Germany}
\author{Mikko M\"{o}tt\"{o}nen}
\email{mikko.mottonen@aalto.fi}
\affiliation{QCD Labs, COMP Centre of Excellence, 
Department of Applied Physics, Aalto University, 
PO Box 13500, FI-00076 Aalto, Finland}

\title[]{Cryogenic microwave source based on nanoscale tunnel junctions}

\begin{document}

%%%%%%%%%%%%%%%%%%%%%%%%%%%%%%%%%%%%%%%%%%%%%%%%%%%%%%%%%%%%%%%%%%%%%
%% The "tocentry" environment can be used to create an entry for the
%% graphical table of contents. It is given here as some journals
%% require that it is printed as part of the abstract page. It will
%% be automatically moved as appropriate.
%%%%%%%%%%%%%%%%%%%%%%%%%%%%%%%%%%%%%%%%%%%%%%%%%%%%%%%%%%%%%%%%%%%%%
%\begin{tocentry}
%\end{tocentry}

%%%%%%%%%%%%%%%%%%%%%%%%%%%%%%%%%%%%%%%%%%%%%%%%%%%%%%%%%%%%%%%%%%%%%
%% The abstract environment will automatically gobble the contents
%% if an abstract is not used by the target journal.
%%%%%%%%%%%%%%%%%%%%%%%%%%%%%%%%%%%%%%%%%%%%%%%%%%%%%%%%%%%%%%%%%%%%%
\begin{abstract}
\begin{spacing}{1.2}
We experimentally realize an incoherent microwave source driven by voltage-controlled quantum tunneling of electrons through nanoscale normal-metal--insulator--superconductor junctions coupled to a resonator.
We observe the direct conversion of the electronic energy into microwave photons by measuring the power spectrum of the  microwave radiation emitted from the resonator.
The demonstrated total output power exceeds that of 2.5-K thermal radiation although the photon and electron reservoirs are at subkelvin temperatures.
Measurements of the output power 
quantitatively agree with a theoretical model in a wide range of the bias voltages providing information on the electrically-controlled photon creation.
The developed photon source is fully compatible with low-temperature electronics and offers convenient in-situ electrical control of the photon emission rate with a predetermined frequency, 
without relying on intrinsic voltage fluctuations of heated normal-metal components nor suffering from unwanted dissipation in room temperature cables.
In addition to its potential applications in microwave photonics, our results provide complementary verification of the working principles of the recently discovered quantum-circuit refrigerator. \\
\end{spacing}
\end{abstract}

\section*{Introduction}
Superconducting circuits provide a promising platform for quantum technological applications \cite{Wolf2009}, such as  quantum information processing \cite{Blais2004,Wallra2004,Majer2007,Sillanpaa2007,Devoret2013,Kelly2015,Ofek2016}, sensing \cite{Inomata2016,Govenius2016,Saira2016}, and refrigeration of  electric components \cite{Clark2005,Giazotto2006,Timofeev2009,Kuan2016}.
% and a fertile ground for study of non-equilibrium thermodynamics and statistical mechanics \cite{Golubev2015}.
%\textcolor{blue}{For efficient implementation of functional devices it is important to understand their interplay with the surrounding microwave radiation \cite{Giazotto2006,Meschke2006,Timofeev2009,Pekola2010,Pascal2011,Jones2012A,Golubev2015,Matti2016}.}
In particular, microwave photons constitute a fundamental and controllable medium \cite{Hofheinz2009} for energy and information transport between quantum electric components.
For example, quantum-limited heat conduction between nanoelectronic components mediated by microwave photons has been measured  \cite{Meschke2006,Timofeev2009}, even through coplanar waveguides (CPW) across macroscopic distances \cite{Matti2016}.

In this context, an on-chip tunable photon source with high dynamic range is desirable since it can provide reference photons with a given frequency for calibration of cryogenic devices such as microwave photon detectors \cite{Govenius2016}. Importantly, the photons created on a chip minimize unwanted and difficult-to-calibrate losses which typically occur in microwave cables and connectors.
Thus, on-demand creation of microwave photons \cite{Houck2007,Peng2016} and improvement of the dynamic range of the sources
%control of resonator states \cite{Hofheinz2009,Hofheinz2011,Srinivasan2014,Pierre2014} 
are fundamental to electric quantum circuits.

%In this context, microwave photons in superconducting resonators and waveguides are routinely used as information and energy carriers between different components\cite{Blais2004,Wallra2004,Majer2007,Sillanpaa2007,Yin2013,Inomata2014,Pechal2014}.  
%The quantum--limited heat transfer between nanoelectronic components mediated by the microwave photons has been measured  \cite{Meschke2006,Timofeev2009}, even through high--quality coplanar waveguides (CPWs) across macroscopic distances \cite{Matti2016}.
%Thus, on-demand control of photon states in microwave resonators \cite{Hofheinz2009,Jones2013a,Srinivasan2014,Pierre2014} and understanding the energy transfer mechanisms based on microwave photons are of great interest for these quantum circuits \cite{Giazotto2006,Meschke2006,Timofeev2009,Pekola2010,Pascal2011,Jones2012A,Golubev2015,Matti2016}. 
It is well known that tunneling of electrons across a nanoscale barrier can lead to energy exchange with the coupled electromagnetic environment such as a resonator\cite{Giazotto2006,Kuan2016,Pekola2010,Devoret1990,Girvin1990,Averin1990,Ingold}. %\cite{Kuan2016,Ingold,Giazotto2006,Timofeev2009,Pekola2010,Matti2016}.
Consequently, several schemes of resonator state control based on the quantum tunneling of electrons have been introduced using, for example, Josephson junctions \cite{Bajjani2010,Hofheinz2011}, superconducting qubits \cite{You2007,Astafiev2007,Hauss2008,Grajcar2008}, and quantum dots \cite{Childress2004,Liu2014,Stockklauser2015,Bruhat2016}. Even refrigeration of a superconducting microwave resonator mode has been demonstrated using controlled tunneling of single electrons across normal-metal--insulator--superconductor (NIS) tunnel junctions\cite{Kuan2016}. Our photon source is based on similar tunneling of electrons across  NIS junctions. It does not require any external magnetic fields to operate and can potentially offer a wide dynamic range and robust control of the output power, in contrast to, for example, the one based on tunneling of Cooper pairs across a Josephson junction~\cite{Hofheinz2011}, the power of which is determined by a sharp resonance condition in its bias voltage. To the best of our knowledge, no microwave photon source based on metallic NIS tunnel junctions has been demonstrated to date.

%It has the advantage (\textcolor{red}{compared to what?}) of all electrical in-situ control of the microwave photon state and potentially can be used to control the electromagnetic environment in low-temperature electronic devices \cite{Jones2013a}. 

%External microwave injection has been used for manipulations of the control of cQED devices. However increase of the injection of microwave with various frequency to integrated devices will be demanding.

In this Letter,  we report the fabrication and operation of a microwave source driven by single-electron tunneling through NIS junctions.
The experimental setup is extended from that of the first quantum-circuit refrigerator\cite{Kuan2016}, such that we can measure the power spectrum of the microwave radiation generated in a CPW resonator.
%In contrast to Ref. \citenum{Kuan2016}, we directly observe the microwave light generated by the electron tunneling process.
%The output-power measurement and the model of the energy transfer quantify the  photon creation controlled by the bias voltage across the SIN junctions.
%The model, which shows  excellent qualitative agreements with the experimental observations, reveals the resonator state under the photon creation operation.
The device is fully compatible with low-temperature electronics and offers orders of magnitude electrical tunability of the output power with the frequency corresponding to the lowest mode of the CPW resonator. 
Our results demonstrate that the effective mode temperature of the resonator can be driven beyond 2.5 K, far above the temperatures of the phonon and electron reservoirs of the system.
Thus our source has potential in delivering relatively high powers without the excess heating of other nearby components.
%compared to the one based on tunneling of Cooper pairs across a Josephson junction that is limited by the critical current.
%In contrast to the previous research where the effective mode temperature of the resonator was estimated to be 60mK \cite{Hofheinz2011}, we have realized more than 2K of the effective mode temperature which is higher than critical temperature of aluminum.

%{\bf Sample and measurement setup.} 
\section*{Results}
\subsection*{Samples and the measurement setup}
The experimental sample shown in Figure~\ref{sample_layout3}a--c consists of NIS tunnel junctions capacitively coupled to a half-wave-length superconducting CPW resonator.
Figure~\ref{sample_layout3}d depicts the layout of the sample and the measurement setup. 
%The coupling capacitances are denoted by $C_{1-4}$.
Microwave photons, which are generated in the resonator at 10-mK phonon temperature, decay to the transmission line which is capacitively coupled to the resonator, and are characterized using a room temperature spectrum analyzer. 
To estimate the output power generated by controlled electron tunneling without off-sets,
we subtract the output power corresponding to zero bias voltage across the superconductor--insulator--normal-metal--insulator--superconductor (SINIS) junction from the output power at finite bias voltages.
As depicted in Figure~\ref{sample_layout3}b, a pair of NIS junctions is used to create the photons, while another pair is used as a thermometer to measure the electron temperature of the normal-metal island \cite{Leivo1996} (see Supporting Information). 

To enhance the interaction between the tunneling electrons and the resonator, we use a large coupling capacitance between the normal-metal island and the resonator, $C_3$, in comparison to the tunnel junction capacitance, $C_\textrm{J}$.
A small enough coupling capacitance between the resonator and the transmission line is employed to maintain 
a well-defined resonance described by
the loaded resonator quality factor $Q\approx 60$.
The length of the resonator is chosen to obtain the desired frequency of the output radiation.
In this letter, we examine two samples with resonator lengths of $L_\textrm{res}=13.14$ mm (Sample A)
and $L_\textrm{res}=6.89$ mm (Sample B) which correspond to the fundamental resonance frequencies of 4.55 GHz and 8.32 GHz, respectively.

\subsection*{Principle of photon generation}
Photon-assisted electron tunneling is the key phenomenon leading to the photon creation in our experiments.
When an electron tunnels across an NIS junction, it can absorb energy from or emit energy to the resonator, that is, annihilate or create photons in the resonator \cite{Ingold}. 
Figure~\ref{sample_layout3}e illustrates single-electron tunneling events across the NIS junctions and the associated photon creation. 
When the energy provided by the bias voltage to the electron tunneling across the junction, $eV_{\rm B}/2$, is larger than
the superconductor gap parameter, $\Delta$, the photon creation rate increases faster as a function of the bias voltage than the annihilation rate. 
Thus, the effective temperature related to the creation and annihilation of photons increases.
%\textcolor{blue}{This should be explained more clearly or simply removed. (We used $U$ for the actual bias voltage across the junction instead of $V_{\rm B}$ because it can be different from the bias voltage in general.)}
%The electron tunneling controlled by the external bias voltage is utilized to drive photon creation in the coplanar waveguide resonator.
The elastic tunneling does not directly affect the resonator modes but dominates the electric current across the junctions for $eV_{\rm B}/(2\Delta)\gtrsim 1$.

%{\bf Results and discussion.}
%We measure the output power from the resonator into the transmission line using a spectrum analyzer as shown in Figure~\ref{sample_layout3}d.
\subsection*{Experiments on Sample A}
Figure~\ref{sum_apple_3_23_16}a,b shows the output power spectrum of our microwave source as a function of the bias voltage for Sample A.
%Panels a and b show the same thing just plotted in a different way.
We observe a peak in the output power density around $4.55$~GHz matching the frequency of the fundamental resonator mode estimated using the experimental parameters given in Table \ref{table_pA}. 
The output power density increases with the bias voltage for $eV_{\rm B}/(2\Delta)>1$ whereas it is almost vanishing for $|eV_{\rm B}/(2\Delta)|<1$. 
This onset of output radiation matching the energy gap in the superconductor density of states provides clear evidence that single-electron tunneling is responsible for the observed photon emission.
The bias-voltage dependence of the frequency-integrated output power density, i.e., the net output power is shown in Figure~\ref{sum_apple_3_23_16}c.
%, where
%the power density in Fig.~\ref{sum_apple_3_23_16}(b) is integrated with respect to the frequency from 4.4 GHz to 4.8 GHz.
%The inset shows the dependence of the output power on the bias voltage in a wider voltage regime.
The onset of positive net power is clearly visible at $eV_{\rm B}/(2\Delta)=1$.

Theoretically, the net output power is related to the average photon numbers of the resonator, $\bar{n}_{\rm res}$, and the transmission line, $\bar{n}_{\rm TL}$, at the fundamental angular frequency, $\omega_0$, by (see Supporting Information)
\begin{eqnarray}
P_\textrm{RT} = \frac{2C_4^2\hbar\omega_0^3Z_0}{L_\textrm{res}c_\textrm{res}}(\bar{n}_{\rm res}-\bar{n}_{\rm TL}),
\label{netpower}
\end{eqnarray}
where $Z_0$ is the characteristic impedance of the transmission line and $c_\textrm{res}$ is the capacitance per unit length of the resonator. 
In our experiments, the resonator mode is in a thermal state, and hence we may describe its average photon numbers using the Bose distribution function as
\begin{eqnarray}
\bar{n}_{\rm res} = \frac{1}{\exp[\hbar\omega_0 /(k_{\rm B}T_{\rm res})]-1}.
\label{numb1}
\end{eqnarray}
With a given $\bar{n}_{\rm TL}$, the above equations provide an analytic way of extracting the resonator temperature, $T_{\rm res}$, and average photon number directly from the output power.

Although we measure only the difference in the output powers at finite and zero bias voltages, we assume for simplicity that the net output power at zero bias voltage vanishes, i.e., here the resonator is thermalized with the tranmission line. This assumption does not lead to changes in the estimated $T_\textrm{res}$, but renders $\bar{n}_\textrm{TL}$ to describe the cumulative voltage-bias-independent heating effect of the resonator owing to the transmission line and any additional environments (see Supporting Information). Consequently, we refer below to $T_\textrm{TL}$ as the apparent transmission line temperature. Note that the estimated resonator temperature becomes insensitive to changes in $T_\textrm{TL}$ at high bias voltages where $\bar{n}_\textrm{res}\gg\bar{n}_\textrm{TL}$.

%the output power can be alternatively expressed using the temperature of the fundamental resonator mode $T_{\rm res}$ because
%the average photon number of the fundamental mode, $\bar{n}_{\rm res}$, is written with $T_{\rm res}$ as 
%\begin{eqnarray}
%\bar{n}_{\rm res} = \frac{1}{\exp[\hbar\omega_0 /(k_{\rm B}T_{\rm res})]-1}.
%\label{numb1}
%\end{eqnarray}

Figure~\ref{sum_apple_3_23_16}d shows the resonator temperature and average photon number corresponding to the measured output power in Figure~\ref{sum_apple_3_23_16}c as functions of the bias voltage.
These results indicate that the mode temperature and the average photon number can be efficiently controlled using the bias voltage. 
For $eV_{\rm B}>10\Delta$, we have $T_\textrm{res}>2.5$~K.
Such a high mode temperature is not conveniently achieved by coupling the resonator to a hot resistor due to the transition of the superconducting aluminum, employed as the lead material, to the normal state.

As mentioned above, the electron temperature at the normal-metal island of the NIS junctions is measured using a pair of NIS junctions (see Supporting Information).
Importantly, the electron temperature shown in Figure~\ref{VT2}b is much lower than that of the resonator mode for almost any bias voltage in Figure~\ref{sum_apple_3_23_16}c,d. 
This observation verifies that the microwave radiation observed by the spectrum analyzer directly arises from photon-assisted electron tunneling rather than from heating of the normal-metal island.

Although negative net output power is challenging to differentiate in the measured power spectra,
we observe in Figure~\ref{sum_apple_3_23_16}c a shallow but statistically significant dip in the experimental output power around $eV_{\rm B}/(2\Delta)=1$.
The dip corresponds to the refrigeration of the fundamental mode owing to photon-assisted tunneling, and hence provides complementary evidence of this phenomenon first observed in Ref.~\citenum{Kuan2016}.
In contrast to our measurements of the output radiation, Ref.~\citenum{Kuan2016} employs a probe resistor to study the resonator temperature.
%Note that the mode temperature is obtained to rise well above the critical temperature of the aluminum for $eV_{\rm B}\gg 2 \Delta$.

%As observed above, the average photon number in the resonator can be increased by the voltage-controlled electron tunneling process. It is also affected by the coupling strength between the resonator and the transmission line. 
\subsection*{Thermal model}
To describe in detail the energy transfer from the tunneling electrons to the transmission line, mediated by microwave photons in the fundamental mode of the resonator, we develop a thermal model shown in Figure~\ref{energy_trans1}.
Here, we consider thermal states for the electric components assuming that the temperatures of the resonator mode, $T_{\rm res}$, and of the electrons in the normal-metal island, $T_{\rm N}$, are well defined in the parameter range studied.
We have verified the validity of this assumption with a model that accounts for transitions between individual resonator states but do not present the model here since it is unnecessarily complicated.
Because of the relatively weak coupling between the resonator and the transmission line, we assume that a tunneling electron does not directly induce photon-assisted transitions in the transmission line. 
Instead, the influence of the tunneling events is indirectly taken into account through the change in the temperature of the resonator mode.

In the thermal model, the power from the tunneling electrons to the resonator, $P_{\rm JR}$, is calculated using the $P(E)$ theory \cite{Ingold}, as detailed in Supporting Information together with the above-discussed power flow from the resonator to the transmission line, $P_{\rm RT}$. 
Both of these powers are functions of the temperature of the resonator mode.
The output power into the transmission line is calculated by finding the mode temperature, at which these powers balance, $P_{\rm JR}=P_{\rm RT}$.
The apparent temperature of the transmission line, corresponding to the photons moving towards the resonator and possible additional heating channels, is assumed to be independent of the applied bias voltage.
This is justified for the photons by the employed circulator shown in Figure~\ref{sample_layout3}d.
The experimentally measured electron temperature of the normal-metal island is used in the simulation.

%As described in detail in Supporting Information, the output power is solved from the thermal model using the parameters given in Table \ref{table_pA}.
The theoretical results, employing the parameters given in Table \ref{table_pA}, show 
good agreement with the experiments in a wide range of output powers (see Figure~\ref{sum_apple_3_23_16}c).
%We set $T_{\rm transmission line}=180$ mK so that the theoretical value of the bias voltage with which the resonator starts to be heated and the output power becomes positive fits to the experimental value.
%\textcolor{red}{Discuss here more throughly how the parameters were obtained.}
In the model, the dip in the output power around $eV_{\rm B}/(2\Delta)=1$ is fully arising from photon absorption induced by electron tunneling.
Furthermore, the modelled average number of photons, $\bar{n}_{\rm res}$, and the temperature of the fundamental mode, $T_{\rm res},$ are shown in Figure~\ref{sum_apple_3_23_16}d as functions of the bias voltage.
%In thermal equilibrium, the mode temperature is related to the average number of photons as 
%\begin{eqnarray}
%\bar{n}_{\rm res} = \frac{1}{\exp[\hbar\omega_0 /(k_{\rm B}T_{\rm res})]-1}.
%\label{numb1}
%\end{eqnarray}
%Figure~\ref{sum_apple_3_23_16}(d) also shows the temperature of the fundamental mode obtained using the experimental results in Fig.~\ref{sum_apple_3_23_16}(c) and Eqs.~(\ref{numbers}) and (\ref{PRT}).

\subsection*{Experiments on Sample B}
Experimental results similar to those discussed above are shown in Figure~\ref{sum_A1_3_23_26} for Sample B.
Figure~\ref{sum_A1_3_23_26}a shows the output power density as a function of the frequency and the bias voltage.
A peak appears around $8.3$ GHz for $eV_{\rm B}/(2\Delta)>1$ and becomes taller with increasing bias voltage.
Figure \ref{sum_A1_3_23_26}b exhibits the traces of the output power density, which are integrated for Figure~\ref{sum_A1_3_23_26}c to obtain the net output power as a function of the bias voltage.
The qualitative behavior of the output power matches that of Sample A:
the output power begins to increase with bias voltage around $eV_{\rm B}/(2\Delta)=1$,
where we observe a shallow dip corresponding to cooling of the resonator.
However, the magnitude of the output power is greater and the mode temperature shown in Figure~\ref{sum_A1_3_23_26}d is lower compared with Figure~\ref{sum_apple_3_23_16}.
This is because the coupling of the resonator to the transmission line is proportional to the third power of the resonance frequency [see Eq.~(\ref{netpower})] which is almost twice as large here in comparison to Sample A.
%Figure~\ref{sum_A1_3_23_26}d represents $T_{\rm res}$ and $\bar{n}_{\rm res}$ corresponding to the measured output power in Figure~\ref{sum_A1_3_23_26}c as functions of bias voltage.
%For $eV_{\rm B}>18\Delta$, $T_{\rm res}$ is higher than 2.5 K.
%Note that the mode temperature is obtained to rise well above the critical temperature of the aluminum for $eV_{\rm B}\gg 2 \Delta$.
%Figure \ref{sum_A1_3_23_26}(d) shows the temperature of the fundamental mode as a function of control voltage calculated using the experimental results and Eqs.~(\ref{numbers}) and (\ref{PRT}).
%The figure also exhibits the average photon number $\bar{n}_{\rm res}$ of the fundamental mode of the resonator 
%and the temperature of the fundamental mode as a function of control voltage obtained from the model with the parameters in Table \ref{table_pA}.
%The output power increases almost linearly for $eV_{\rm B}/(2\Delta)>1.6$.
%The solid curve is the numerical results based on the thermal model.
%Again, the theoretical results well agree with the experimental observations in a wide range of the bias voltage.
%We set $T_{\rm transmission line}=150$ mK.
%Figure \ref{sum_A1_3_23_26}(c) shows the modelled 
% $T_{\rm res}$ and $\bar{n}_{\rm res}$ as functions of the control voltage.
% Figure4

\section*{Discussion}
In conclusion, we have measured the spectrum of microwave radiation generated by photon-assisted electron tunneling through NIS junctions.
The implemented microwave source realizes direct conversion of the electrostatic energy provided by the dc voltage source to microwave photons. 
Importantly, it
does not rely on the thermal voltage fluctuation spectrum of a resistor, and hence may provide higher effective temperatures, faster tunability, and less excess heating than usual thermal methods.
Namely, our results demonstrate the control of the output power up to the regime where the effective mode temperature of the resonator is higher than the critical temperature of the widely used aluminum.
The device is compatible with low-temperature electronics and offers in-situ electrical tunability of the output power with the frequency predetermined by the associated resonator. 
Our measurements of the output power are in good agreement with the corresponding theoretical model, 
heavily suggesting that our interpretation that photon-assisted tunneling is mainly responsible for the output power is correct.
This interpretation is verified by the observation of much higher output powers than what is expected for resonator temperatures matching the normal-metal electron temperature.

\section*{Methods}
\subsection*{Sample fabrication}
The experimental sample shown in Figure~\ref{sample_layout3}a--c is fabricated on a high-purity 500-$\mu$m-thick silicon wafer. 
The silicon substrate is passivated with a 300-nm-thick thermally grown silicon dioxide.
The half-wave-length superconducting CPW resonator is defined with photolithography and reactive ion etching of a 200-nm-thick sputtered Nb layer.
Subsequently, a 50-nm-thick layer of Al$_2$O$_3$ is introduced on the whole wafer using atomic layer deposition (ALD) process at 200 $^\circ$C. This ALD oxide serves as the dielectric material of the parallel plate capacitors $C_{1-4}$ shown in Figure~\ref{sample_layout3}d. 
The NIS junctions and a transmission line connected to the resonator are subsequently defined using electron beam lithography,
followed by a standard two-angle evaporation with in-situ oxidation to form the tunnel barriers. The final NIS nano-structure consists of a 20-nm-thick normal metal (Cu) on top of a 20-nm-thick superconductor (Al), separated by a thin aluminum-oxide tunnel barrier.

%\section*{ASSOCIATED CONTENT}
\subsection*{Supporting Information}
%\subsection*{Supporting Information}
The Supporting Information is available free of charge.
Parameters, Apparent transmission line temperature, Energy transfer between tunneling electron and resonator,
Energy transfer between resonator and transmission line, Thermometer. %(PDF)
%\section*{AUTHOR INFORMATION}
\subsection*{Corresponding Authors}
*E-mail: shumpei.masuda@aalto.fi
*E-mail: mikko.mottonen@aalto.fi

\subsection*{Author Contributions}
Authors S.M. and K.Y.T fabricated the samples and conducted the experiments. 
S.M. also developed the model and analysed the data.
M.P. contributed to the sample fabrication, measurements, and data analysis. 
R.L. and J.G. contributed to the measurements.
M.S and H.G. contributed to the theoretical analysis.
M.M. provided the initial ideas and suggestions for the experiment, and supervised the work in all respects.
All authors commented on the manuscript written by S.M. and M.M.

\subsection*{Competing Financial Interest}
The authors declare no competing financial interest.

\begin{acknowledgement}
We acknowledge the provision of facilities and technical support by Aalto University at OtaNano - 
Micronova Nanofabrication Centre. We have received funding from the European
Research Council under Starting Independent Researcher Grant No.~278117
(SINGLEOUT) and under Consolidator Grant No.~681311 (QUESS), the Academy of Finland through its Centres of Excellence Program
(project nos 251748 and 284621) and grants (Nos. 265675, 286215,
276528, 305237, and 305306), the Emil Aaltonen Foundation, the Jenny and Antti Wihuri Foundation, the Alfred Kordelin Foundation, and the
Finnish Cultural Foundation.
We thank Leif Gr\"{o}nberg for assistance in sample fabrication.
%We thank H. Grabert and  M. Silveri for useful discussions.
\end{acknowledgement}

\section*{Figures and tables}
\begin{figure}
\begin{center}
\includegraphics[width=15cm]{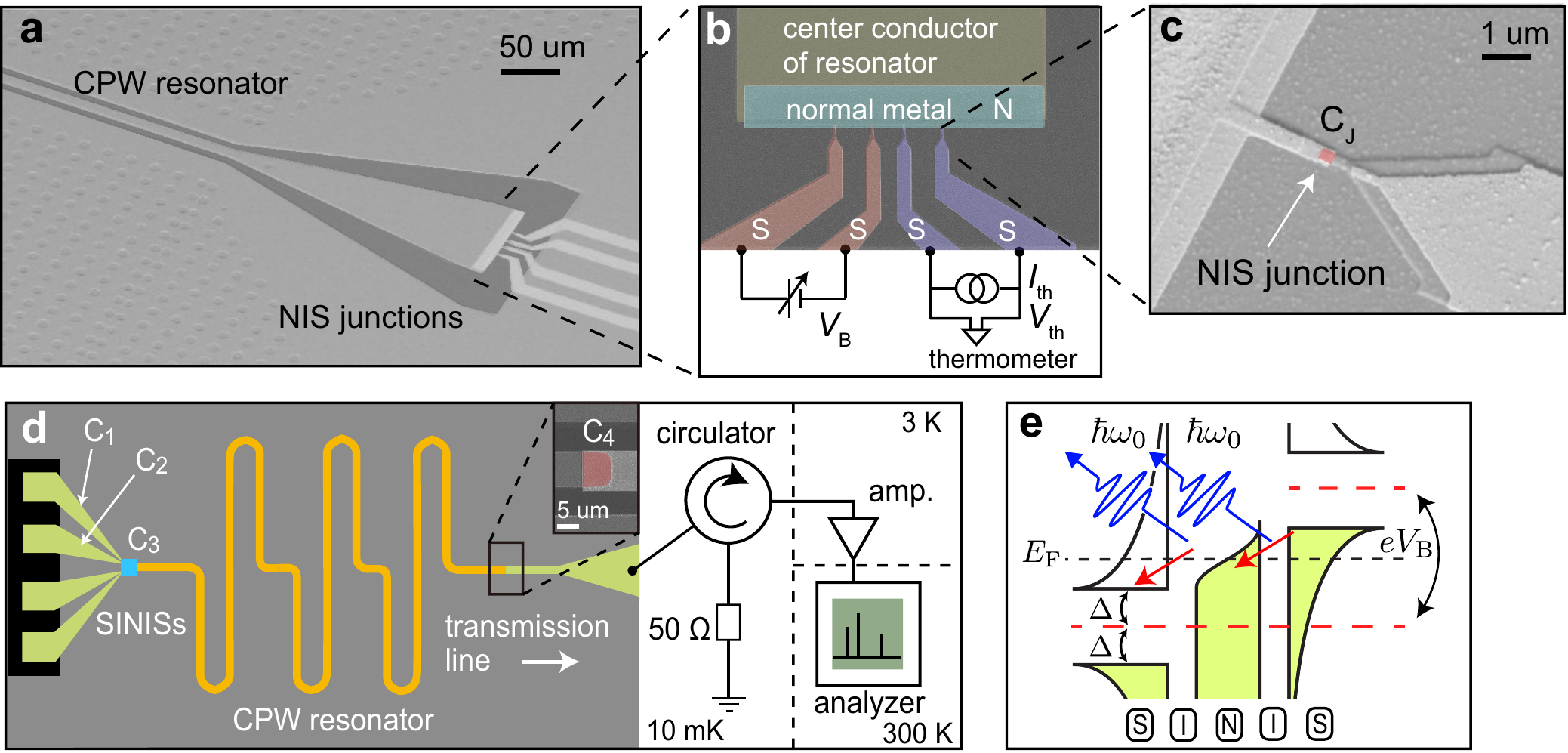}
\end{center}
  \caption{(a) Scanning-electron microscope (SEM) image of a fabricated device illustrating the CPW resonator and the NIS junctions operating together as the microwave source. (b) Colored SEM image of the NIS junctions. The junctions highlighted in red are used to excite the resonator using the bias voltage $V_{\rm B}$. The junctions highlighted in blue are used as a thermometer to measure the electron temperature of the normal-metal island.
(c) SEM image of an NIS junction highlighted in red. 
%The tunnel junction area is approximately 200$\times$200~nm$^2$ inferred from the SEM image of the sample.
(d) Device design and measurement scheme. 
Here, $C_{1-2}$, $C_{3}$, and $C_4$ denote the coupling capacitances 
between the bonding pads and the ground plane, between the normal-metal island of the NIS junctions and the center conductor of the resonator, and between the resonator and the transmission line (see inset), respectively, and
$C_{\rm J}$ denotes the junction capacitance.
The area indicated by the blue square corresponds to pannel~(a) although the axis of view is different. 
%The inset shows the region indicated by the black square, where the center conductor of the resonator is forming parallel-plate capacitor $C_4$ highlighted in red with the center conductor of the transmission line.  
(e) Energy diagram for photon-assisted single-electron tunneling at bias voltage $eV_{\rm B}/(2\Delta)>1$, where $\Delta$ is the superconductor gap parameter and $\hbar\omega_0$ is the energy of an emitted photon.
The black solid curves at the normal metal and the superconductors represent the Fermi--Dirac distribution function and the density of states in the superconductors, respectively. The colored areas represent the occupied states.}
\label{sample_layout3}
\end{figure}

\begin{figure}
\begin{center}
\includegraphics[width=13cm]{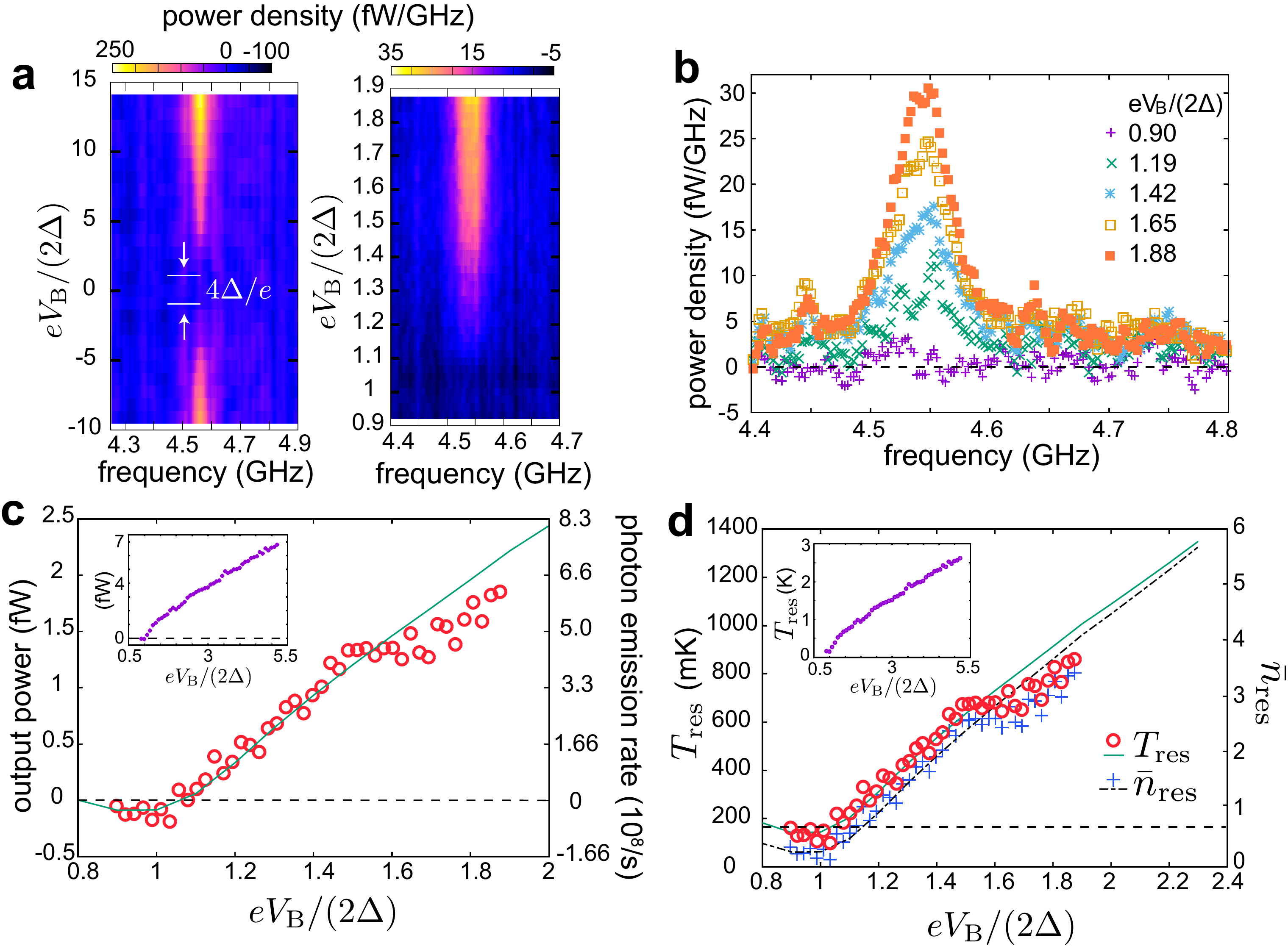}
\end{center}
\caption{
Experimental and numerical results for Sample A with the fundamental resonance frequency $f_0=4.55$ GHz. 
(a) Resonator output power densities measured as functions of frequency and bias voltage. Each frequency trace is measured by a spectrum analyzer as shown schematically in Figure~\ref{sample_layout3}d and averaged over 21000 repetitions.
In addition, we employ a three-point moving average in frequency.
The left panel shows the output power spectrum in a wider voltage range compared with the right panel.
In contrast to Figure~\ref{sample_layout3}d, no circulator was used for the data in the left panel. 
(b) Output power densities as functions of frequency for the indicated bias voltages.
The horizontal dashed line indicates the zero level corresponding to zero bias voltage.
(c) Output power obtained by integrating data as in (b) from 4.4 GHz to 4.8 GHz as a function of the bias voltage.  The green curve indicates the theoretical prediction of the thermal model illustrated in Figure~\ref{energy_trans1}.
The inset shows the output power for an extended bias voltage range.
The horizontal dashed lines indicate the zero level.
(d) Average photon number (blue crosses) and temperature (red circles) of the fundamental mode as functions of the bias voltage obtained using the experimental  results in (c) and Eqs.~(\ref{netpower}) and (\ref{numb1}). The parameters are given in Table \ref{table_pA}.
Here, the dashed and solid curves represent $\bar{n}_{\rm res}$ and $T_{\rm res}$
obtained from the thermal model, respectively.
We used a 38.5-dB cryogenic amplifier, 24-dB room temperature amplifier, and 6-dB room temperature attenuator.
We assumed 15.9 dB of loss from the PCB and rf cables as opposed to 13.3 dB measured using a control sample.
The horizontal dashed line indicates the apparent temperature of the transmission line as given in Table \ref{table_pA}.
}
\label{sum_apple_3_23_16}
\end{figure}

\begin{figure}
\begin{center}
\includegraphics[width=7cm]{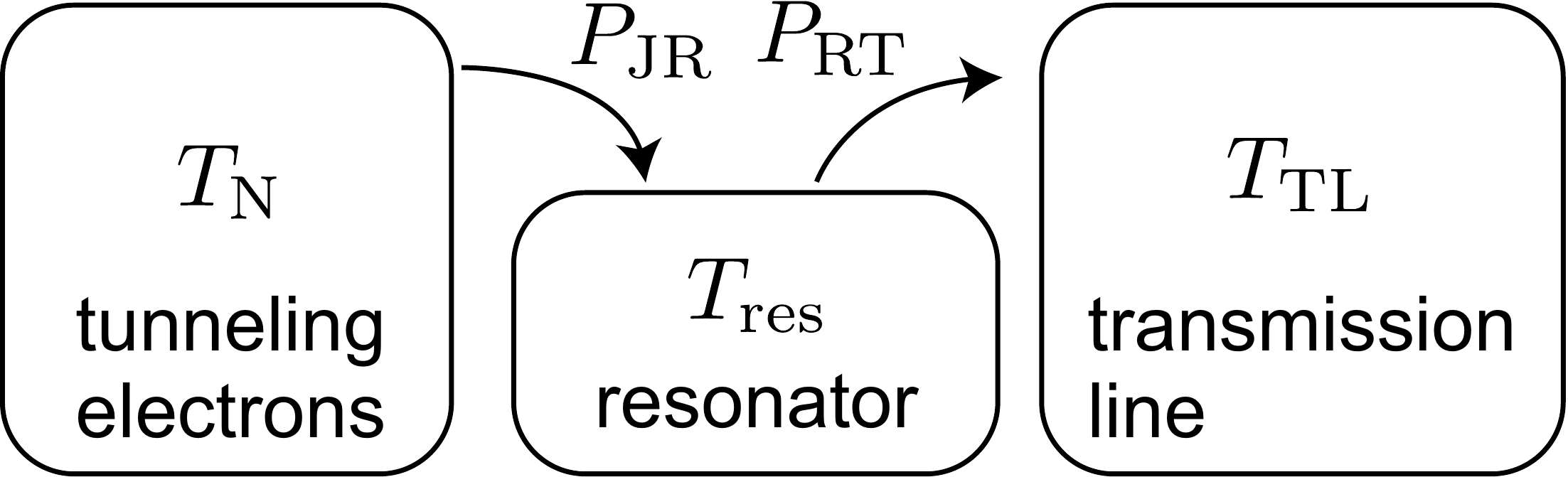}
\end{center}
\caption{
Thermal model of the device indicating the relevant temperatures and energy flows. 
The temperature of the fundamental resonator mode, $T_{\rm res}$, becomes higher than the apparent temperature of the transmission line, $T_{\rm TL}$, when the applied bias voltage strongly drives the photon creation.
Here, $T_{\rm N}$ is the electron temperature at the normal-metal island of the NIS junctions.
At steady state, the power $P_{\rm JR}$ from the tunneling electrons to the resonator and the power $P_{\rm RT}$ from the resonator to the transmission line balance, $P_{\rm JR}=P_{\rm RT}$.
See Supporting Information for details.}
\label{energy_trans1}
\end{figure}

\begin{figure}
\begin{center}
\includegraphics[width=13cm]{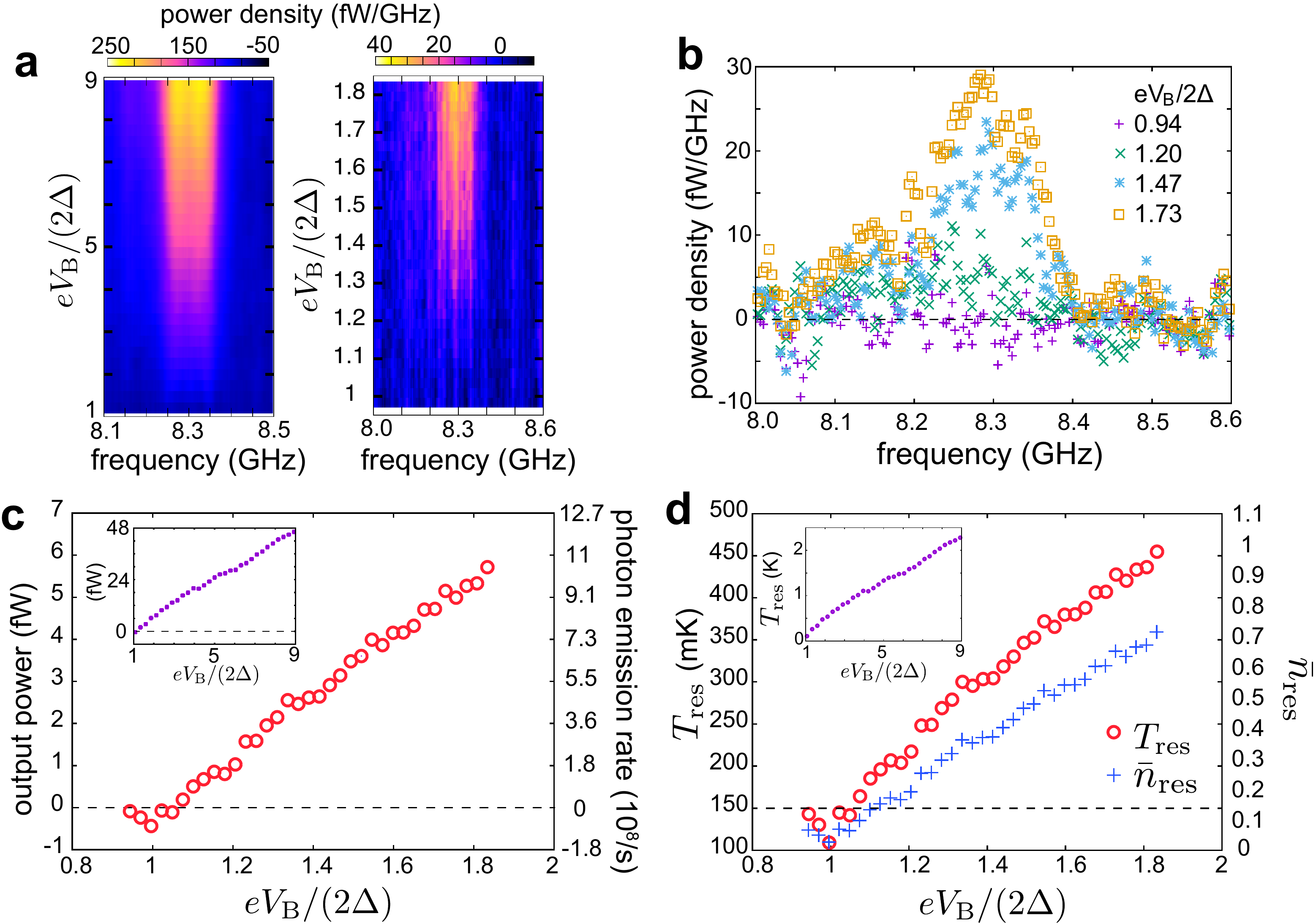}
\end{center}
\caption{
Experimental results for Sample B with the fundamental resonance frequency $f_0=8.3$~GHz. 
(a)~Resonator output power densities measured as functions of frequency and bias voltage. 
Each frequency trace is measured by a spectrum analyzer as shown schematically in Figure~\ref{sample_layout3}d and averaged over 21000 repetitions. In addition, we employ a three-point moving average in frequency.
The left panel shows the output power spectrum in a wider voltage range compared with the right panel.
%In contrast to Figure \ref{sum_apple_3_23_16}a for Sample A, the left and the right panels show the results for $eV_{\rm B}/(2\Delta)>1$.
(b) Output power densities as functions of frequency for the indicated bias voltages.
The horizontal dashed line indicates the zero level.
(c) Output power obtained by integrating data as in (b) from 8.0 GHz to 8.6 GHz as a function of the bias voltage.  
The inset shows the output power for an extended bias voltage range.
The horizontal dashed lines indicate the zero levels.
(d) Average photon number (blue crosses) and temperature (red circles) of the fundamental mode as functions of the bias voltage obtained using the data in (c) and Eqs.~(\ref{netpower}) and (\ref{numb1}). The parameters are given in Table \ref{table_pA}.
We used a 40-dB cryogenic amplifier, 24-dB room temperature amplifier, and 6-dB room temperature attenuator.
We assumed 22 dB of loss from the PCB and rf cables as opposed to 19.5 dB measured using a control sample.
The horizontal dashed line indicates the apparent temperature of the transmission line as given in Table \ref{table_pA}.
In contrast to Figure~\ref{sum_apple_3_23_16}c,d, we do not show theoretical results here due to the lack of reliable temperature data for the normal-metal island.}
\label{sum_A1_3_23_26}
\end{figure}

\clearpage

\begin{table}
  \caption{
 Parameters of the experimental samples: the length of the resonator $L_{\rm res}$, superconductor gap parameter $\Delta$,
  Dynes parameter $\gamma_{\rm D}$, capacitance per unit length $c_{\rm res}$,
  inductance per unit length $l_{\rm res}$, fundamental resonance frequency $f_0$, tunnel resistance $R_{\rm T}$, junction capacitance $C_{\rm J}$, capacitance $C_{3/4}$, 
  characteristic impedance of the transmission line $Z_0$, apparent temperature of transmission line $T_{\rm TL}$, the employed  attenuation of the PCB and rf cables $\alpha$, the width of the center conductor $w$, and the separation between the center conductor and the ground plane of the CPW resonator $s$. 
%We extracted $\Delta$ from the NIS current-voltage characteristics. 
Capacitance $C_{1/2}$ is more than 50 times larger than $C_3$.}
  \label{tbl:example}
  \begin{tabular}{c|c|c||c|c|c|}
& SAMPLE A & SAMPLE B & &SAMPLE A & SAMPLE B \\
\hline
$L_{\rm res}$ & 13.14 mm & 6.89 mm & $C_3$ & 0.84 pF & 0.78 pF \\
\hline
$\Delta$ & 220 $\mu$eV & 191 $\mu$eV & $C_4$ & 0.072 pF & 0.079 pF \\
\hline
$\gamma_{\rm D}$ & $4\times 10^{-4}$ & $4\times 10^{-4}$  & $Z_0$ & 53 $\Omega$ & 52 $\Omega$ \\
\hline
 $c_{\rm res}$ & 159 pF/m & 169 pF/m  & $T_{\rm TL}$ & 180 mK & 150 mK \\
 \hline
 $l_{\rm res}$ & 0.45 $\mu$H/m & 0.45 $\mu$H/m & $\alpha$ & 15.9 dB $^{\rm a}$ & 22 dB $^{\rm b}$ \\
 \hline
 $f_0$ & 4.55 GHz & 8.32 GHz & $w$ & 7.8 $\mu$m & 8.0 $\mu$m  \\
 \hline
 $R_{\rm T}$ & 12.5 k$\Omega$ & 10.0 k$\Omega$ & $s$ & 5.7 $\mu$m & 6.1 $\mu$m \\
 \hline
 $C_{\rm J}$ & 3 fF & 3 fF & &  & \\
 \hline
\end{tabular}
 \\

$^{\rm a,b}$ The values experimentally measured with a control sample are 13.3 dB$^{\rm a}$ and 
 19.5 dB$^{\rm b}$.
  \label{table_pA}
\end{table}

\clearpage

\setcounter{figure}{0}
\renewcommand*{\thefigure}{S\arabic{figure}}
\setcounter{table}{0}
\renewcommand*{\thetable}{S\arabic{table}}
\setcounter{page}{1}
\setcounter{equation}{0}
\renewcommand*{\theequation}{S\arabic{equation}}

\begin{center}
\section{Supporting information:\\
Cryogenic microwave source based on nanoscale tunnel junctions}
Shumpei Masuda, Kuan Y. Tan, Matti Partanen, Russell E. Lake, Joonas Govenius, Matti~Silveri, Hermann Grabert and Mikko M\"{o}tt\"{o}nen
\end{center}

\subsection{Parameters}
The fundamental resonance frequency of a homogeneous half-wavelength CPW resonator of length $L_{\rm res}$ is given by
\begin{eqnarray}
f_0=\frac{c}{2L_{\rm res}\sqrt{\varepsilon_{\rm eff}}}
\label{eqf0}
\end{eqnarray}
where the speed of light in vacuum is denoted by $c$ and the effective relative permittivity of the resonator by $\varepsilon_{\rm eff}$.
The CPW capacitance $c_{\rm res}$ and inductance $l_{\rm res}$ per unit length can be expressed as \cite{Gevorgian1995,Goppl2008}
\begin{eqnarray}
c_{\rm res} &=& 4\varepsilon_0\varepsilon_{\rm eff}\frac{K(k_0)}{K(k_0')}, \nonumber\\
l_{\rm res} &=& \frac{\mu_0 K(k_0')}{4K(k_0)},
\label{eqClLl}
\end{eqnarray}
where $K$ is the complete elliptic integral of the first kind and
\begin{eqnarray}
k_0 &=& \frac{w}{w+2s},\nonumber\\
k_0' &=& \sqrt{1-k_0^2},
\label{eqk0}
\end{eqnarray}
depend on the width of the center conductor $w$, and on the separation between the center conductor and the ground plane of the CPW resonator $s$.
We calculate $c_{\rm res}$ and $l_{\rm res}$ in Eq.~(\ref{eqClLl}) using Eqs.~(\ref{eqf0}) and~(\ref{eqk0}) and the measured values for $f_0$, $L_{\rm res}$, $w$, and $s$ given in Table \ref{table_pA}.
We neglect the kinetic inductance because it is sufficiently smaller than the geometric inductance in Eq.~(\ref{eqClLl}) \cite{Goppl2008}, although in principle, the total inductance per unit length is the sum of the geometric and the kinetic contributions.
The resonance frequency calculated using Eq.~(\ref{eqf0}) and the analytic form of $\varepsilon_{\rm eff}$ for an unshielded CPW\cite{Gevorgian1995} has less than 1\% difference from the measured value for Sample A and about 4\% deviation for Sample B.
The characteristic impedance $Z_{0}$ of the transmission line is obtained from $Z_0=\sqrt{l_{\rm res}/c_{\rm res}}$.

Capacitances $C_k$ for $k\in\{1,2,3,4\}$ are estimated using 
\begin{eqnarray}
C_k = \frac{A_k\varepsilon}{d},
\label{eqC}
\end{eqnarray}
where $A_k$ is the area of the parallel-plate capacitor, the permittivity of the aluminum oxide dielectric is denoted by $\varepsilon\approx9.8\times\varepsilon_0$, where $\varepsilon_0$ is the permittivity of vacuum,
and $d=50$~nm is the thickness of the aluminum oxide layer.
The capacitance density used for $C_{\rm J}$ is 75~fF/$\mu$m$^2$.
The tunnel junction area is approximately 200$\times$200~nm$^2$ inferred from the SEM image of the sample.

In the absence of an electromagnetic environment, the current through an NIS junction is given by
\begin{eqnarray}
I(V_{\rm B}) &=& \frac{1}{eR_{\rm T}}\int_{-\infty}^\infty {\rm d}E\Big{[}n_{\rm S}(E)f(E,T_{\rm S})[1-f(E+eV_{\rm B}/2,T_{\rm N})] 
\nonumber\\
&& - n_{\rm S}(E-eV_{\rm B}/2)[1-f(E-eV_{\rm B}/2,T_{\rm S})] f(E,T_{\rm N})\Big{]}, 
\label{eqI3}
\end{eqnarray}
where $T_{\rm N}$ is the electron temperature of the normal-metal island obtained using the thermometer junctions.
%as explained in \ref{Thermomitor}.
The superconductor density of states $n_{\rm S}$ is represented by
\begin{eqnarray}
n_{\rm S}(E) = \Big{|}\mbox{Re}\Big{[}\frac{E/\Delta+i\gamma_{\rm D}}{\sqrt{(E/\Delta+i\gamma_{\rm D})^2 - 1}}\Big{]} \Big{|},
\label{eqns}
\end{eqnarray}
where $\gamma_{\rm D}$ is the Dynes parameter \cite{Giazotto2006} and
$f$ is the Fermi--Dirac distribution function
\begin{eqnarray}
f(E,T) = \frac{1}{e^{E/(k_{\rm B}T)}+1}.
\end{eqnarray}
The tunnel resistance $R_{\rm T}$, the Dynes parameter, and the superconductor gap parameter $\Delta$ are obtained from a fit of Eq.~(\ref{eqI3}) to measured junction $IV$ characteristics.

\subsection{Apparent transmission line temperature}

In our thermal model in Figure~\ref{energy_trans1}, the resonator is thermally connected only to the transmission line and to the tunneling electrons. Due to the simple way the transmission line heats the resonator given by Eq.~\eqref{netpower} however, any additional constant heating power to the resonator is accounted in this model by an adjustment of the average photon number of the transmission line. This gives rise to the apparent temperature of the transmission line, $T_\textrm{TL}$, connected to the average photon number, $\bar{n}_\textrm{TL}$, through the Bose distribution function. Note that $\bar{n}_\textrm{TL}$ is independent of the bias voltage since also the electron temperature of the 50-$\Omega$ resistor in Figure~\ref{sample_layout3}d giving rise to the photons incident on the resonator does not change with the bias voltage.

Since the power arising from the photon-assisted tunneling is negligible at zero bias voltage, we have $P_\textrm{RT}=0$ in Eq.~\eqref{netpower}, and hence $T_\textrm{res}=T_\textrm{TL}$ at zero bias. Thus if we find $T_\textrm{TL}$ at any bias voltage, we obtain the temperature and the average photon number of the resonator at zero bias. Consequently, we obtain these quantities at any bias voltage using Eq.~\eqref{netpower} and the directly measured power change between the finite and zero bias points.

We use the bias point, at which the measured power change crosses zero, as the calibration voltage for $T_\textrm{TL}$. Here, the net output power  given by Eq.~\eqref{netpower} and hence the temperature of the resonator coincides with that at zero bias. Thus the power arising from photon-assisted tunneling, $P_\textrm{JR}$, must vanish. As we show in the next section, this calibration point is independent of many of the sample parameters such as the tunnel resistance, $R_\textrm{T}$. It is also independent of the gain of our amplification chain. Thus our model accurately yields the resonator temperature at the calibration point which equals the apparent transmission line temperature.

%Dynes parameter $\gamma_{\rm D}$ in Eq.~(\ref{eqns}) was extracted from measured current bias curve.

\subsection{Energy transfer between tunneling electrons and the resonator}
\label{Energy transfer between tunneling electron and resonator}
The electron tunneling may lead to creation and annihilation of photons at the resonator
giving rise to the power transfer $P_{\rm JR}$ in Figure~\ref{energy_trans1}.
% using the $P(E)$ function.
%We first calculate the probability $P(E)$ that energy $E$ is deposited to the environment when an electron tunnels, and then introduce the modulated $P(E)$ below.
To this end, we employ the $P(E)$ theory which has been developed to obtain the electron tunneling rates across tunnel junctions and the resulting energy transfer to the environment \cite{Ingold}. 
In terms of the $P(E)$ theory, the environment refers to the electrical degrees of freedom which are coupled to the tunneling process. 

\begin{figure}
\begin{center}
\includegraphics[width=11cm]{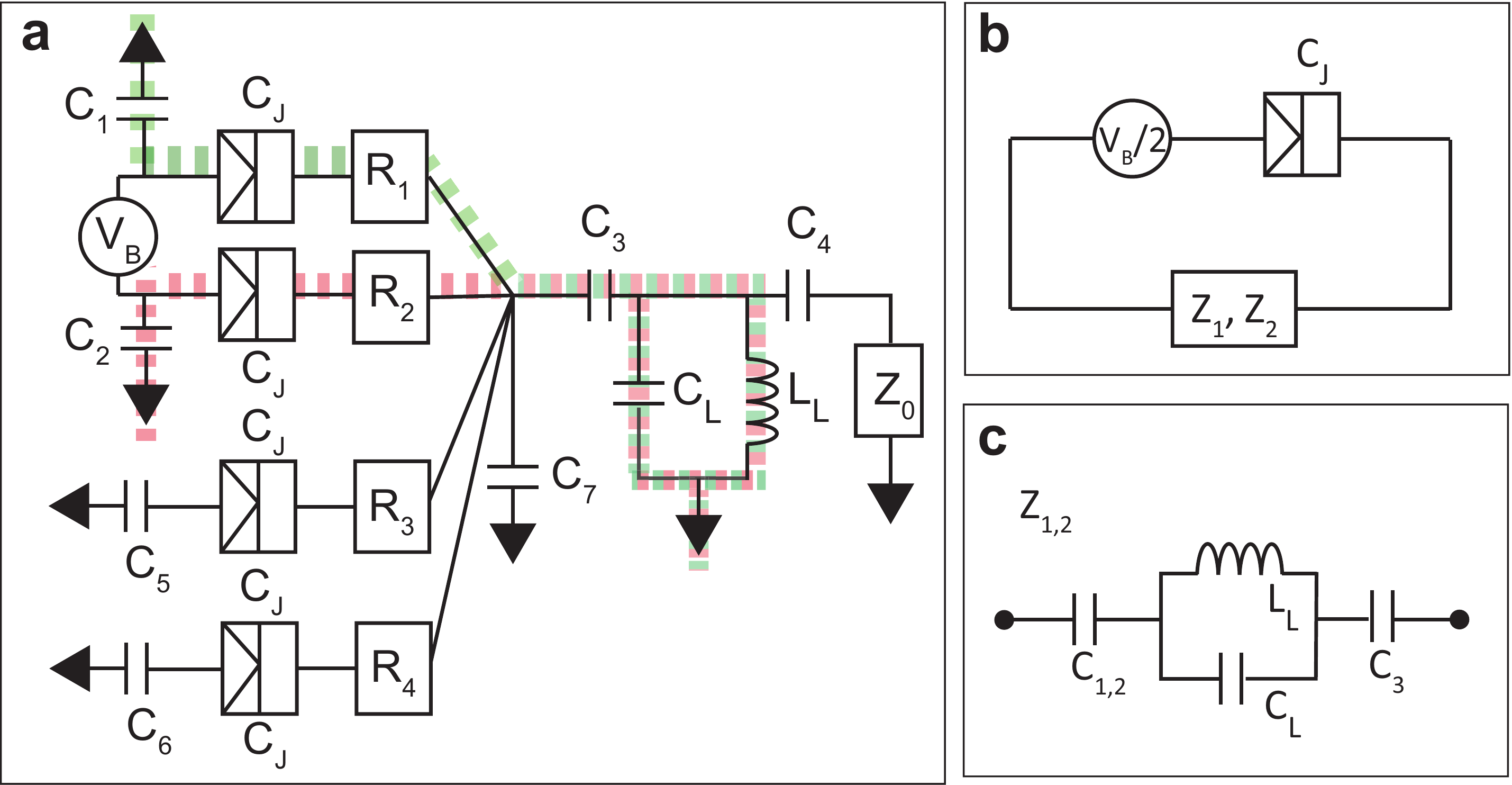}
\end{center}
\caption{(a) Circuit model of the realized microwave source. The resonator is modelled by a single parallel $LC$ circuit.
See Figure 1 for the definition of the capacitances. 
The resistances arise from the finite conductance of the normal-metal island of the NIS junctions.
(b) Effective circuits for photon-assisted tunneling at each NIS junction. The systems with $Z_{1}$ and $Z_{2}$ correspond to the red and green  conduction paths in panel (a), respectively. 
(c) Composition of the impedance elements $Z_1$ and $Z_2$. Here, the finite conductance of the normal-metal island and the direct interplay between the electron tunneling and losses to the transmission line described by $Z_0$ have been neglected.
}
\label{circuit_example2_1}
\end{figure}

The power transfer to the environment strongly depends on the electrical impedance of the system.
A simplified electrical circuit diagram of our system is depicted in Figure~\ref{circuit_example2_1}a.
The first harmonic mode of the CPW resonator is modeled as a parallel $LC$ circuit with capacitance $C_{\rm L}=c_{\rm res}L_{\rm res}/2$, inductance $L_{\rm L}=2l_{\rm res}L_{\rm res}/\pi^2$ and angular frequency $\omega_0=1/\sqrt{C_{\rm L}L_{\rm L}}$. 
We assume that the electron tunneling event on either of the junctions is hardly affected by another junction because of a small junction capacitance $C_{\rm J}$ and large tunneling resistance $R_{\rm T}$, and that the direct influence of the transmission line to the electron tunneling is neglected due to a relatively small coupling capacitance $C_4$.
Thus, we consider a pair of separate effective circuits depicted in Figures~\ref{circuit_example2_1}b,c to calculate the tunneling rate at each NIS junction.
We assume that the resistance of the normal-metal island is small enough to be neglected because it is in series with the small junction capacitance, and hence negligible current flows through it with respect to Ohmic losses.
%In $P(E)$ theory, the effective impedance is represented as parallel 
%We consider the case the junction capacitances $C_{\rm J}$ and the capacitances $C_{4/7}$ are sufficiently smaller than $C_{\rm L}$ and $C_{1/2/3}$ so that
%the influence of the resistance $R_{1/2/3/4}$ and the capacitance $C_{4/7/{\rm J}}$ to the effective impedance related to %electron tunneling is minor compared with the other electrical elements.
%Furthermore, we assume that the electron tunneling event on either of the junctions is hardly affected by another junction because of small $C_{\rm J}$ and large tunneling resistance $R_{\rm T}$,
%and that the direct influence of $Z_0$ to the electron tunneling is neglected due to small $C_4$.
% and is included effectively via temperature of the resonator $T_{\rm res}$ later.

The power from the tunneling electrons to the environment is represented by
\begin{eqnarray}
P_{\rm env}(V_{\rm B}/2,T_{\rm S}, T_{\rm N}) = 2[\overset{\rightarrow}{P}(V_{\rm B}/2,T_{\rm S}, T_{\rm N}) + \overset{\leftarrow}{P}(V_{\rm B}/2,T_{\rm S}, T_{\rm N})],
\label{PJC}
\end{eqnarray}
where the powers caused by the forward and the backward electron tunneling events at each junction are given by \cite{Ingold}
\begin{eqnarray}
\overset{\rightarrow}{P}(V_{\rm B}/2,T_{\rm S}, T_{\rm N}) &=& \frac{1}{e^2R_{\rm T}}\iint_{-\infty}^\infty {\rm d}E{\rm d}E' n_{\rm S}(E)f(E,T_{\rm S})[1-f(E'+eV_{\rm B}/2,T_{\rm N})] \nonumber\\
&&\times(E-E')P(E-E',T_{\rm env}),\nonumber\\
\overset{\leftarrow}{P}(V_{\rm B}/2,T_{\rm S}, T_{\rm N}) &=& \frac{1}{e^2R_{\rm T}}\iint_{-\infty}^\infty {\rm d}E{\rm d}E' n_{\rm S}(E'-eV_{\rm B}/2)[1-f(E'-eV_{\rm B}/2,T_{\rm S})] f(E,T_{\rm N})\nonumber\\
&&\times(E-E')P(E-E',T_{\rm env}),
\label{power1}
\end{eqnarray}
respectively.
Here, $T_{\rm S}$ and $T_{\rm env}$ denote the temperatures of the superconductor and of the environment, respectively. 
%The influence of the finite tunnelling probability of the junction is effectively taken into account via junction resistant $R_{\rm T}$ in Eq. (\ref{power1}).
The factor of two in Eq.~(\ref{PJC}) arises from the fact that the two effective circuits in Figure~\ref{circuit_example2_1}b
have an identical contribution to the power.
%Here, $P(E)$ is the probability of the external electric circuit to absorb the energy $E$ in the electron tunneling process.
The probability density $P(E)$ for the environment to absorb energy $E$ during a tunneling event is expressed as \cite{Ingold}
\begin{eqnarray}
P(E,T_{\rm env}) = \frac{1}{2\pi\hbar}\int_{-\infty}^{\infty} {\rm d}t \exp[J(t,T_{\rm env})]\exp\Big{[} i\frac{E}{\hbar}t \Big{]},
\label{PE1}
\end{eqnarray}
where
\begin{eqnarray}
J(t,T_{\rm env}) &=& 2\int_0^\infty \frac{{\rm d}\omega}{\omega}\frac{\mbox{Re}[Z_{\rm t}(\omega)]}{R_{\rm K}}
\Big{[}\coth\Big{(}\frac{\hbar\omega}{2k_{\rm B}T_{\rm env}}\Big{)} [\cos(\omega t)-1] - i\sin(\omega t)\Big{]},
\label{eqJ}
\end{eqnarray}
and
\begin{eqnarray}
\mbox{Re}[Z_{\rm t}] &=& \frac{\pi}{2C_{\rm L}}\big{[} \delta(\omega-\omega_{0}) + \delta(\omega+\omega_{0})\big{]}
+ \frac{\pi}{2}\frac{C_{1(2)}+C_3}{C_{1(2)}C_3}\delta(\omega).
\label{eqZ}
\end{eqnarray}
In $P(E)$ theory, the effective impedance $Z_{\rm t}$ in Eq.~(\ref{eqJ}) is the impedance of the external circuit $Z_{1/2}$ in parallel with $C_{\rm J}$.
The influence of $C_{\rm J}$, which is approximately 3 fF $\ll C_{1/2/3}$, to the effective impedance is neglected in Eq.~(\ref{eqZ}) because $Z_{\rm t}$ is dominated by $Z_{\rm 1/2}$.
%Finite $C_{1/2/3}$ broadens the peaks of $P(E)$ which become delta functions in the limit of infinite $C_{1/2/3}$.
The influence of $C_{1/2}$ is also negligible because it is more than 50 times larger than $C_3$ and approximately cancel out in Eq.~(\ref{eqZ}).
% approximated $Z_{\rm t}(=1/[i\omega C_{\rm J} + Z_{1(2)}^{-1}])$ by $Z_{1(2)}(=Z_{2})$
%Finite $C_{1/2/3}$ induces a zero-energy peak to  $P(E)$ which is otherwise a superposition of delta functions at positive and negative resonator frequencies.

\begin{figure}
\begin{center}
\includegraphics[width=14cm]{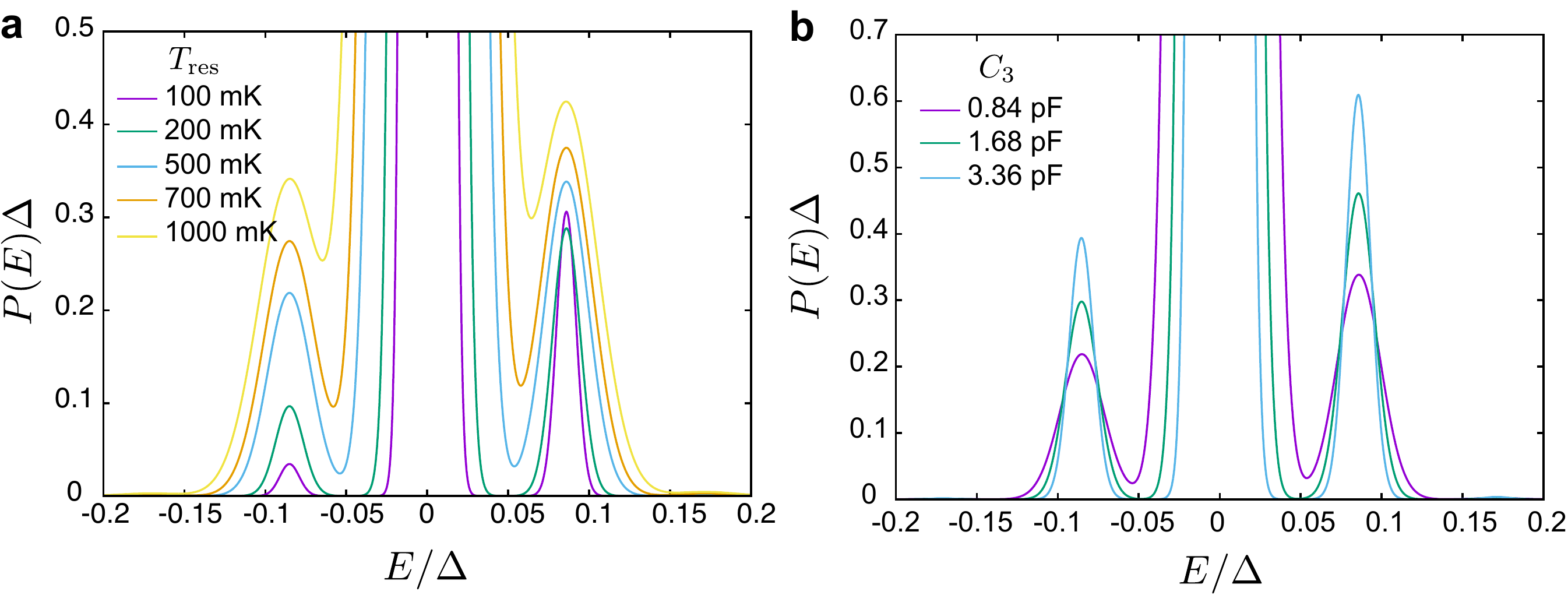}
\end{center}
\caption{
(a) $P(E)$ function of Sample A for $eV_{\rm B}/(2\Delta) = 1.2$ and $T_{\rm res}$ as indicated.
The other parameters are given in Table \ref{table_pA}.
(b) $P(E)$ of Sample A  shown for various indicated $C_3$, for $T_{\rm res}=500$ mK and $eV/(2\Delta) = 1.2$.
 }
\label{PE2}
\end{figure}

Note that the environment consists not only of the resonator but also of the capacitors $C_{1/2/3}$.
Finite $C_{1/2/3}$ broadens the peaks of $P(E)$ which become delta functions in the limit of infinite $C_{1/2/3}$.
%To mimic this partial energy transfer we assume that the emitted or absorbed energy $E$ by the tunneling electrons can contribute to $P_{\rm JR}$ if the energy $E$ is sufficiently close to the energy corresponding to the resonance frequency, $\pm \hbar\omega_0$; on the other hand, if the frequency corresponding to the energy $E$ is off-resonant, the excitation can not be transferred to the transmission line; and the energy dissipates to somewhere else. 
%As explained in Sec. \ref{Another model of energy transfer between resonator and transmission line} $P_{\rm RT}$ originates  the voltage at the capacitor $C_4$. The photon transfer to the transmission line is diminished for off-resonant excitation because the voltage at the resonator fluctuates.
%To mimic the energy transfer we introduce a modulation function $g_{\rm M}(E)$ which consists of two Gaussian functions located at $\pm \hbar\omega_0$ with the unity peak-amplitude and the width corresponding to the FWHM of the peaks of the power density experimentally obtained and displayed in Figs.~\ref{sum_apple_3_23_16}(a) and \ref{sum_A1_3_23_26}(a) for sample A and B, respectively. 
%The modulation function is defined by
%\begin{eqnarray}
%g_{\rm M}(E) = 
%\exp\Big{[}-\frac{(E-\hbar \omega_0)^2}{\Delta_{\rm M}^2}\Big{]} + %\exp\Big{[}-\frac{(E+\hbar \omega_0)^2}{\Delta_{\rm M}^2}\Big{]}
%\end{eqnarray}
%with
%\begin{eqnarray}
%\Delta_{\rm M} = \frac{h {\rm FWHM}}{2(\ln 2)^{1/2}},
%\end{eqnarray}
%where $h$ is the the Planck constant.
%The modulated $P(E)$ is defined by
%\begin{eqnarray}
%P_{\rm M}(E) = g_{\rm M}(E) P(E).
%\end{eqnarray}
Figure~\ref{PE2}a shows $P(E)$ of Sample A for various $T_{\rm res}$ at $eV_{\rm B}/(2\Delta) = 1.2$, where we assumed $T_{\rm env}=T_{\rm res}$.
The peaks at $E=\hbar\omega_0\approx 0.09\times\Delta$ and $E=-\hbar\omega_0\approx -0.09\times\Delta$ correspond to the emission and absorption of a single photon from the resonator, respectively.
The high peak at $E=0$ corresponds to the elastic tunnelling.
%Similar results are obtained for Sample B in Fig.~\ref{PE2}(b).
%The emission and absorption peaks appear at
%$E=\hbar\omega_0\simeq 0.19\Delta$ and $E=-\hbar\omega_0\simeq -0.19\Delta$, respectively. 
We assume that only the peaks at $|E|=\pm\hbar\omega_0$ contribute to $P_{\rm JR}$. 
Thus, only a part of $P_{\rm env}$ contributes to $P_{\rm JR}$.
We compute $P_{\rm JR}$ using Eqs. (\ref{PJC}) and (\ref{power1}) where we first replace $P$ by %$P_{\rm M}$ defined as 
\begin{eqnarray}
P_{\rm M}(E,T_{\rm res}) &=& P(E,T_{\rm res})  \ \ {\rm for} \ \ |E|\ge\hbar\omega_0/2, \nonumber\\
&=& 0 \ \ \hspace{1.6cm} {\rm for} \ \ |E|<\hbar\omega_0/2.
\label{PE1}
\end{eqnarray}

As shown in Figure~\ref{PE2}a, the peaks become broader with increasing $T_{\rm res}$.
The center peak and the side peaks are not well separated for high $T_{\rm res}$. Thus, the numerical results of Sample A may have considerable error beyond $eV/(2\Delta)=1.4$ corresponding to $T_{\rm res} > 500$~mK [see Figure~\ref{sum_apple_3_23_16}d]. The minor discrepancy between the numerical and experimental results in Figure~\ref{sum_apple_3_23_16}c is attributed to this fact.
%On the other hand, the center peak and the side peaks are well separated for sample B even at $T_{\rm res}\sim 1000$ mK,
%and the numerical and experimental results match also for high bias voltages as shown in Fig.~\ref{sum_A1_3_23_26}(b).
Figure~\ref{PE2}b shows the dependence of $P(E)$ on the coupling capacitance $C_3$.
The peak becomes narrower for larger $C_3$ due to the increasing coupling strength between a tunneling electron and the resonator.

\subsection{Energy transfer between the resonator and the transmission line}
\label{Another model of energy transfer between resonator and transmission line}
\begin{figure}
\begin{center}
\includegraphics[width=7cm]{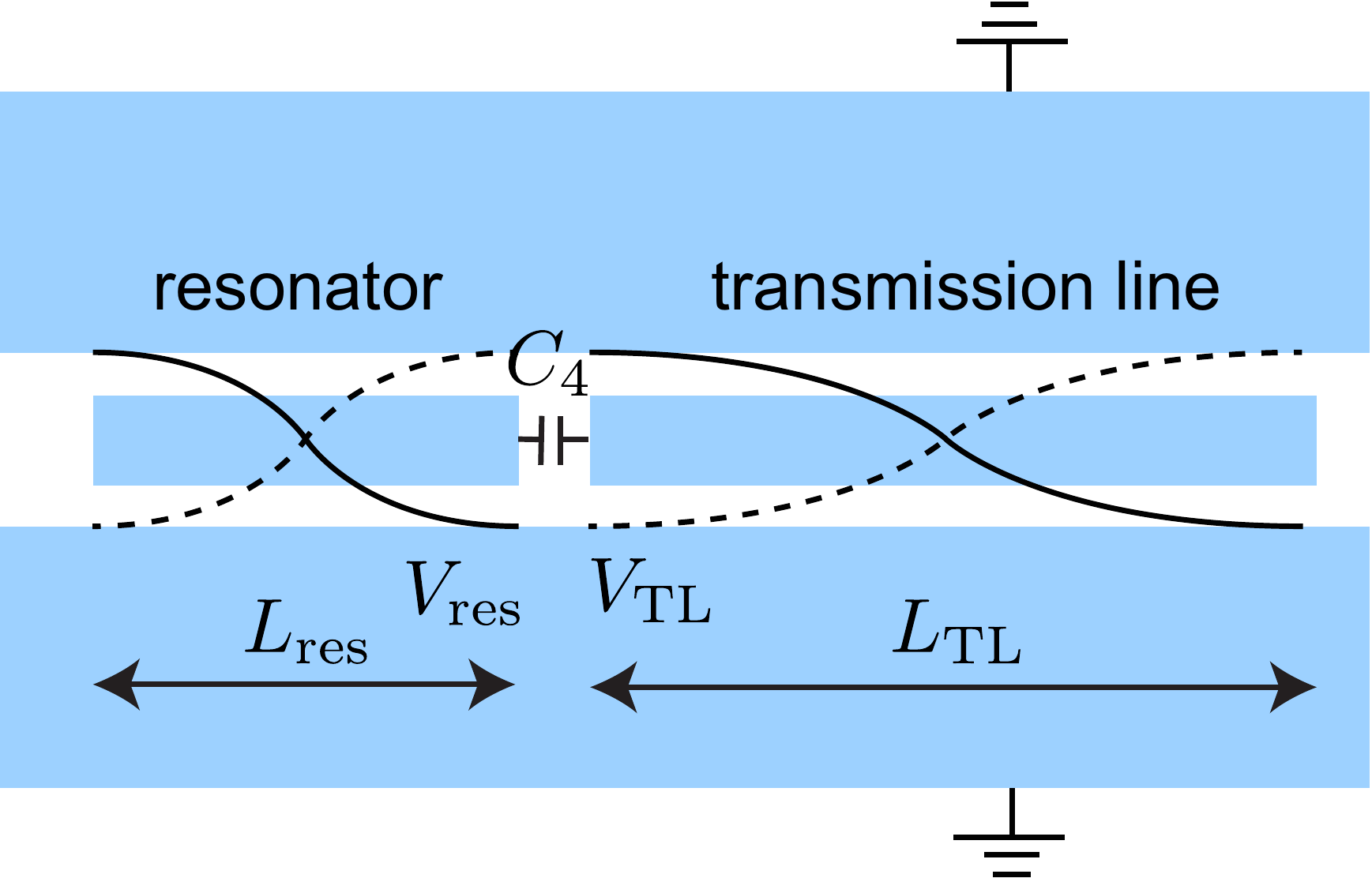}
\end{center}
\caption{
Illustration of a CPW resonator and a transmission line coupled by capacitance $C_4$.
The solid and the dashed lines represent the voltage difference between the center conductor and the ground plane.
The transmission line is modelled by a CPW resonator with length $L_{\rm TL}$ which is eventually extended to infinity.
}
\label{interaction_TR}
\end{figure}

Let us study a model for the energy transfer between a CPW resonator and a transmission line which are mutually  capacitively coupled as shown in Figure~\ref{interaction_TR}.
First, we model the transmission line as a CPW resonator of finite length and finally extend the length to infinity.
The electrostatic energy of the coupling capacitor $C_4$ is given by the voltage difference between the resonator and the transmission line as $C_4(V_{\rm TL} - V_{\rm res})^2/2$, where $V_{\rm res}$ and $V_{\rm TL}$ are the voltage of the resonator and the transmission line at the capacitor, respectively.
Thus, the interaction Hamiltonian between the resonator and the transmission line is represented as
\begin{eqnarray}
\hat{H}_{\rm int} = - C_4 \hat{V}_{\rm res} \hat{V}_{\rm TL}.
\label{Hint1}
\end{eqnarray}
The voltage operators of the resonator $\hat{V}_{\rm res}$ and of the transmission line $\hat{V}_{\rm TL}$
are represented as $\hat{V}_{\rm res}=\sum_M V_M^{\rm (res)}(\hat{a}_M^\dagger + \hat{a}_M)$ and $\hat{V}_{\rm TL}=\sum_{N} V_{N}^{\rm (TL)}(\hat{b}_{N}^\dagger + \hat{b}_{N})$, respectively,
with the creation operator $\hat{a}_M^\dagger$ and the annihilation operator $\hat{a}_M$ of a photon in the $M$th mode of the resonator and the creation operator $\hat{b}_N^\dagger$ and the annihilation operator $\hat{b}_N$ of a photon in the $N$th mode of the transmission line \cite{Blais2004}. The coefficients are written as
$V_M^{\rm (res)}=\sqrt{M\hbar\omega_{\rm res}/(L_{\rm res}c_{\rm res})}$ and $V_{N}^{\rm (TL)}=\sqrt{N\hbar \omega_{\rm TL}/(L_{\rm TL}c_{\rm TL})}$ with the fundamental resonance frequencies $\omega_{\rm res / TL} = \pi/(L_{\rm res / TL)}\sqrt{l_{\rm res / TL }c_{\rm res / TL}})$. 
The length of the resonator (transmission line) is $L_{\rm res}$ ($L_{\rm TL}$), the capacitance per unit length 
is $c_{\rm res}$ ($c_{\rm TL}$) and the inductance per unit length is $l_{\rm res}$ ($l_{\rm TL}$).
Above, we have assumed that $C_4$ is so small that its effect on $\omega_{\rm res / TL}$ and the operators may be neglected.

We employ a basis composed of the energy eigenstates of an uncoupled resonator--transmission-line system.
Assuming weak coupling, we may estimate the transition rate from an energy eigenstate $|i\rangle=|m_1, m_2, \cdots; n_1, n_2, \cdots\rangle$ to
a different one $|f\rangle=|m'_1, m'_2, \cdots; n'_1, n'_2, \cdots\rangle$ using Fermi's golden rule, where $m_M$ ($n_{N}$) denotes the number of photons in the $M$th ($N$th) mode of the resonator (transmission line).
The Hamiltonian in Eq. (\ref{Hint1}) is rewritten as
\begin{eqnarray}
\hat{H}_{\rm int}^{\rm eff} = -C_4\sum_M\sum_{N}V_M^{\rm (res)}V_{N}^{\rm (TL)}
(\hat{a}_M^\dagger \hat{b}_{N} +  \hat{a}_M \hat{b}_{N}^\dagger),
\label{Heff10}
\end{eqnarray}
where we omitted the terms proportional to $\hat{a}_M^\dagger \hat{b}_{N}^\dagger$ and $\hat{a}_M \hat{b}_{N}$ which do not contribute to the transition rates in Fermi's golden rule.
%This is equivalent to the rotating-wave approximation.
We first consider the case where a photon in the $I$th mode of the resonator is annihilated and a photon in the $J$th mode of the transmission line is created, i.e., we only consider the term $ -C_4V_I^{\rm (res)}V_{J}^{\rm (TL)}\hat{a}_I \hat{b}_{J}^\dagger$ in Eq.~(\ref{Heff10}).
%We calculate the probability $P_{if}^{(+)}(t)$ that a photon of $I$th mode is annihilated in the resonator and a photon of $J$th mode is created in the transmission line.
Consequently, the probability $P_{if}(t)$ that we find the state $|f\rangle$ at time $t$ starting from $|i\rangle$ is given by  Fermi's golden rule as
%\begin{eqnarray}
%P_{if}(t) &=& \frac{2\pi}{\hbar}C_4^2 (V_I^{\rm (res)})^2(V_{J}^{\rm (TL)})^2 t \Big\{  \delta[\hbar(J\Omega - I\omega)] \delta_{m_I', m_I-1}\delta_{n_J', n_J+1}\delta_{m_K',m_K}\delta_{n_K',n_K} \nonumber\\
%&&+ \delta[\hbar(-J\Omega + I\omega)]\delta_{m_I', m_I+1}\delta_{n_J', n_J-1}\delta_{m_K',m_K}\delta_{n_K',n_K} \big{)} \Big\},
%\end{eqnarray}
\begin{eqnarray}
P_{if}^{(IJ)}(t) &=& t\frac{2\pi}{\hbar}C_4^2 (V_I^{\rm (res)})^2(V_{J}^{\rm (TL)})^2 m_I(n_J+1)  \nonumber\\ 
&&\times\delta(E_f - E_i) 
\delta_{m_I',m_I-1} \delta_{n_J',n_J+1} \Big{(}\prod_{K\ne I}\delta_{m_K',m_K}\Big{)} \Big{(}\prod_{K\ne J}\delta_{n_K',n_K}\Big{)},
\label{Pif}
\end{eqnarray}
where $E_f - E_i = \hbar(\omega^{(J)}_{\rm TL} - \omega^{(I)}_{\rm res})$ with
$\omega^{(J)}_{\rm TL}=J\omega_{\rm TL}$ and $\omega^{(I)}_{\rm res}=I\omega_{\rm res}$ .
%The probability $P_{if}(t)$ is independent of number of photons if $m_I \ge 1$.

Let us calculate the probability $P_{I}^{(+)}(t)$ that a photon in the $I$th mode of the resonator is annihilated and a photon in the transmission line is created.
To this end, we sum up the probabilities for all photon numbers and every mode $J$ in the transmission line.
First, we execute the summation over the final photon numbers, which results in
\begin{eqnarray}
P_i^{(IJ)} = t\frac{2\pi}{\hbar}C_4^2 (V_I^{\rm (res)})^2(V_{J}^{\rm (TL)})^2 m_I(n_J+1)  \delta[\hbar(\omega_{\rm TL}^{(J)}- \omega_{\rm res}^{(I)})].
\label{PiIJ}
\end{eqnarray}
Next we sum with respect to $J$ in the limit where $L_{\rm TL}$ extends to infinity. 
Thus, we replace the discrete summation over $J$ by an integral with respect to energy using the relation
\begin{eqnarray}
\lim_{L_{\rm TL}\rightarrow \infty} \frac{1}{\Delta E_J}\sum_J \Delta E_J = \lim_{L_{\rm TL}\rightarrow \infty} \frac{1}{\Delta E_J} \int_0^\infty {\rm d}E_J,
\label{sumoverJ}
\end{eqnarray}
where $\Delta E_J=\hbar\omega_{\rm TL}=\hbar\pi/(L_{\rm TL}\sqrt{l_{\rm TL}c_{\rm TL}})$.
Using Eq.~(\ref{sumoverJ}) to sum over $J$ in Eq.~(\ref{PiIJ}), the probability of a photon in the $I$th mode to decay given the initial state $|i\rangle$ is represented by
\begin{eqnarray}
P_i^{(I)}(t) &=& \lim_{L_{\rm TL}\rightarrow \infty}\sum_J P_{i}^{(IJ)}(t) \nonumber\\
&=& \frac{2L_{\rm TL}\sqrt{l_{\rm TL}c_{\rm TL}}}{\hbar^2}C_4^2 (V_I^{\rm (res)})^2(V_{J}^{\rm (TL)})^2 m_\omega(n_\omega+1)t
\nonumber\\
&=& 2C_4^2\frac{\omega^2Z_0}{L_{\rm res}c_{\rm res}}m_\omega(n_\omega+1)t
\label{PEbar}
\end{eqnarray}
where $Z_0 = \sqrt{l_{\rm TL}/c_{\rm TL}}$ and we have introduced a notation $m_\omega := m_I$ and 
$n_\omega$ is the photon number in the initial state of the transmission line corresponding to the mode with angular frequency $\omega := \omega_{\rm res}^{(I)}$.
%$L_{\rm TL}$ was cancelled out and $\omega^{(J)}_{\rm TL}= \omega^{(I)}_{\rm res}$ was used because of $\delta(E_f-E_i)$ in Eq.~(\ref{Pif}).
%For simplicity of notation we replace $n_J$ by $n_I$ hereafter.

Considering all the possible initial states with different photon numbers,
the probability of annihilating a photon in the resonator mode $I$, ${P}_I^{(+)}$, may be represented as
\begin{eqnarray}
{P}_\omega^{(+)}(t)&:=&{P}_I^{(+)}(t) =  
\sum_{m_\omega}\sum_{n_\omega} p_{m_\omega} p_{n_\omega} P_i^{(I)}(t)
\nonumber\\
&=&  2C_4^2\frac{\omega^2Z_0}{L_{\rm res}c_{\rm res}}\bar{m}_\omega(\bar{n}_\omega+1)t,
\end{eqnarray}
where the average photon numbers with angular frequency $\omega$ in the resonator and the transmission line are denoted by $\bar{m}_\omega$ and $\bar{n}_\omega$, respectively.
Here, $p_{m_\omega}$ $(p_{n_\omega})$ is the probability that there are $m_\omega$ $(n_\omega)$ photons with angular frequency $\omega$
in the resonator (transmission line) in the initial state.
Hereafter, we replace ${P}_I^{(+)}$ by ${P}_\omega^{(+)}$ for simplicity of notation.
Thus, the rate that photons in the resonator with the energy $\hbar\omega$ transfer to the transmission line is obtained as
\begin{eqnarray}
\Gamma_{\rm res,TL}^{(\omega)} = \frac{{\rm d}{P}_\omega^{(+)}}{{\rm d}t}  
= 2C_4^2\frac{\omega^2Z_0}{L_{\rm res}c_{\rm res}} \bar{m}_\omega(\bar{n}_\omega+1).
\end{eqnarray}
%where we assumed that the probability that there are $m$ photons of $I$th mode in the resonator is given by
%$P_{I,m}^{(\rm res)} = \exp[-E^{(\rm res)}_{I,m}/(k_B T_{\rm res})]/Z_I^{(\rm res)}$ with
%$Z_I^{(\rm res)} = \sum_{m} \exp[-E^{(\rm res)}_{I,m} / (k_B T_{\rm res})]$ and
%$E^{(\rm res)}_{I,m}= \hbar  \omega^{(I)}_{\rm res}(m + 1/2)$, and 
%the probability that there are $n$ photons of $J$th mode in the TL is given by
%$P_{J,n}^{(\rm TL)} = \exp[-E^{(\rm TL)}_{J,n}/(k_B T_{\rm TL})]/Z_J^{(\rm TL)}$ 
%with $Z_J^{(\rm TL)} = \sum_{n} \exp[-E^{(\rm TL)}_{J,n} / (k_B T_{\rm TL})]$ and
%$E^{(\rm TL)}_{J,m}= \hbar \omega^{(J)}_{\rm TL}(n + 1/2)$.
%Note that here the $J$th mode photon in the transmission line has the same energy as the $I$th mode photon in the resonator.  
In the same manner the photon transition rate from the transmission line to the resonator is obtained as
\begin{eqnarray}
\Gamma_{\rm TL,res}^{(\omega)} =  2C_4^2\frac{\omega^2Z_0}{L_{\rm res}c_{\rm res}} (\bar{m}_\omega+1)\bar{n}_\omega.\label{power2}
\end{eqnarray}
The total power from the resonator to the transmission line mediated by the photons with angular frequency $\omega$ is represented with the photon numbers in the resonator and the transmission line as
\begin{eqnarray}
P_{\rm RT}^{(\omega)}=\hbar\omega [\Gamma_{\rm res,TL} - \Gamma_{\rm TL,res}] &=&  2C_4^2\frac{\hbar\omega^3Z_0}{L_{\rm res}c_{\rm res}} (\bar{m}_\omega - \bar{n}_\omega)\nonumber\\
&=&  C_4^2\frac{\hbar\omega^3Z_0}{C_{\rm L}} (\bar{m}_\omega - \bar{n}_\omega).
\label{PRT}
\end{eqnarray}
In the thermal state, we have
\begin{eqnarray}
\bar{m}_\omega = \frac{1}{e^{\hbar \omega /(k_{\rm B}T_{\rm res})}-1},\nonumber\\
\bar{n}_\omega = \frac{1}{e^{\hbar \omega /(k_{\rm B}T_{\rm TL})}-1},
\label{numbers}
\end{eqnarray}
and hence
\begin{eqnarray}
P_{\rm RT}^{(\omega)} %&=&  2C_4^2\frac{\hbar\omega^3Z_0}{L_{\rm res}c_{\rm res}} 
%\Big{[}
%\frac{1}{e^{\hbar \omega/(k_{\rm B}T_{\rm res})}-1}
%-
%\frac{1}{e^{\hbar \omega/(k_{\rm B}T_{\rm TL})}-1}
%\Big{]}\nonumber\\
&=& C_4^2\frac{\hbar\omega^3Z_0}{C_{\rm L}}
\Big{[}
\frac{1}{e^{\hbar \omega/(k_{\rm B}T_{\rm res})}-1}
-
\frac{1}{e^{\hbar \omega/(k_{\rm B}T_{\rm TL})}-1}
\Big{]},
\label{PRT2}
\end{eqnarray}
owing to Eq.~(\ref{PRT}).

\subsection{Thermometry}
\label{Thermomitor}
A pair of current-biased NIS junctions is used as a thermometer as illustrated in Figure~\ref{sample_layout3}b.
The voltage $V_{\rm th}$ across the thermometer junctions provides a good measure of the electron temperature
$T_{\rm N}$ of the normal-metal island in the temperature range of interest to us.
In $P(E)$ theory, the tunnel current is written as \cite{Ingold}
\begin{eqnarray}
I_{\rm th}(V_{\rm th},T_{\rm N},T_{\rm S},T_{\rm env}) &=& \frac{1}{eR_{\rm T}}\iint_{-\infty}^\infty {\rm d}E{\rm d}E' \Big{[}n_{\rm S}(E)f(E,T_{\rm S})[1-f(E'+eV_{\rm th}/2,T_{\rm N})] 
\nonumber\\
&& - n_{\rm S}(E'-eV_{\rm th}/2)[1-f(E'-eV_{\rm th}/2,T_{\rm S})] f(E,T_{\rm N})\Big{]} P(E-E',T_{\rm env}).\nonumber\\
\label{eqI1}
\end{eqnarray}
The current in Eq.~(\ref{eqI1}) is approximated by
\begin{eqnarray}
I_{\rm th}(V_{\rm th},T_{\rm N}) &=& \frac{1}{eR_{\rm T}}\int_{-\infty}^\infty {\rm d}E\Big{[}n_{\rm S}(E)f(E,T_{\rm S})[1-f(E+eV_{\rm th}/2,T_{\rm N})] 
\nonumber\\
&& - n_{\rm S}(E-eV_{\rm th}/2)[1-f(E-eV_{\rm th}/2,T_{\rm S})] f(E,T_{\rm N})\Big{]}, 
\label{eqI2}
\end{eqnarray}
where we replaced $P(E-E',T_{\rm env})$ by $\delta(E-E')$ because in our case, the current is dominated by the elastic electron tunneling.
Here, we neglect the dependence of $I_{\rm th}$ on $T_{\rm S}$ justified by $k_{\rm B}T_{\rm S}\ll \Delta$.
Thus, the voltage $V_{\rm th}$ essentially depends only on $I_{\rm th}$ and $T_{\rm N}$.
For a fixed value of $I_{\rm th}$, $V_{\rm th}$ can be regarded as a single-valued function of $T_{\rm N}$.
We utilize this property to convert the measured $V_{\rm th}$ into $T_{\rm N}$.

\begin{figure}[h!]
\begin{center}
\includegraphics[width=14cm]{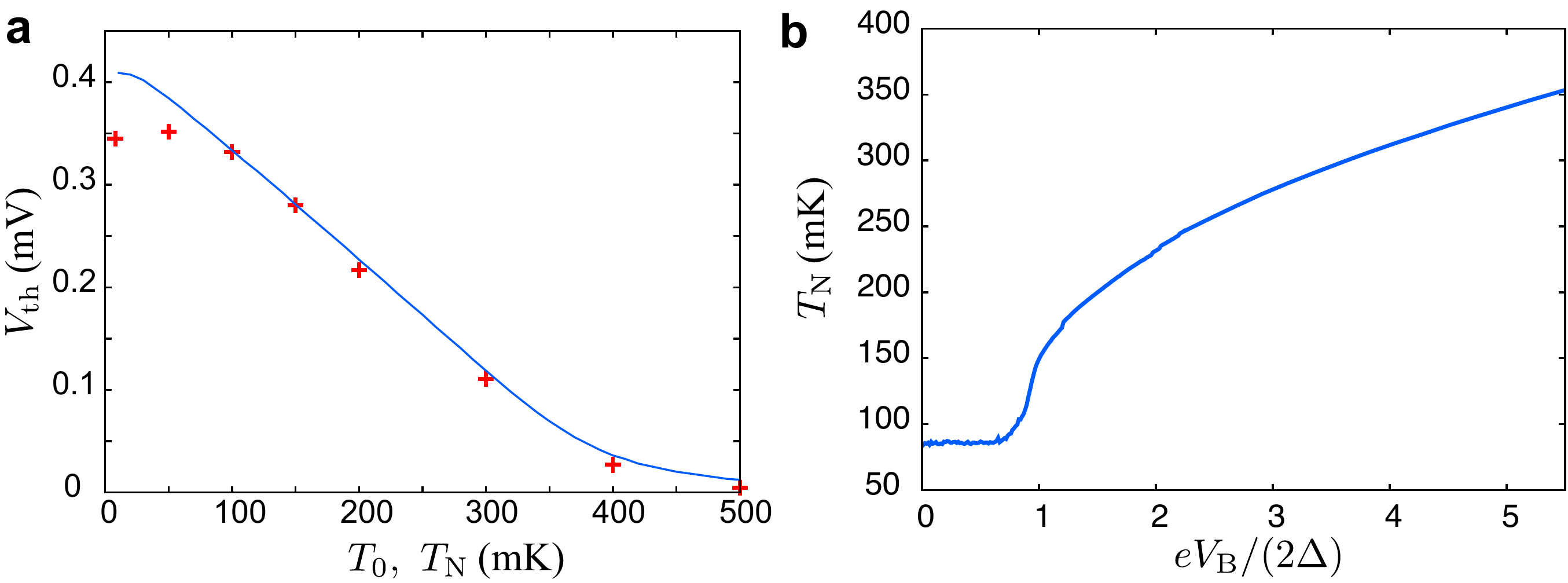}
\end{center}
\caption{
%Thermometer calibration and the bias voltage dependence of the temperature of the normal--metal island. 
(a) Measured voltage (markers) $V_{\rm th}$ across the two thermometer junctions of Sample~A as a function of the phonon bath temperature $T_0$ for $V_{\rm B}=0$ and $I_{\rm th}=$17.7 pA. 
The solid line shows $V_{\rm th}$ as a function of the electron temperature $T_{\rm N}$ as
calculated using Eq.~(\ref{eqI2}).
This result serves as the conversion function from $V_{\rm th}$ into $T_{\rm N}$.
(b) Measured electron temperature as a function of the bias voltage $V_{\rm B}$ in the case of Figure \ref{sum_apple_3_23_16}c,d. 
}
\label{VT2}
\end{figure}

Figure~\ref{VT2}a shows the thermometer voltage as a function of the bath temperature $T_0$ measured for $I_{\rm th}= $17.7 pA and vanishing bias voltage at the other junctions, $V_{\rm B}=0$. 
%The tunnel resistance and superconductor gap parameter of the thermometer junctions are set so that the theoretical result fit to the measured result, and coincide to the ones of the junctions used as photon source.
%It also shows $T_{\rm N}$-dependence of $V_{\rm th}$ calculated using Eq.~(\ref{eqI2}). 
The experimental value of $V_{\rm th}(T_{\rm 0})$ 
matches the theoretical value obtained from Eq.~(\ref{eqI2}) for
$V_{\rm th}(T_{\rm N}=T_{\rm 0})$ and $T_0\gtrsim$ 100 mK.
This is because the electron temperature of the normal-metal island is close to the bath temperature for high $T_0$.
The agreement of the theoretical and numerical results shows that this model of the thermometer is accurate.
On the other hand, the experimental values of $V_{\rm th}$ are lower than the theoretical ones at $T_0 < 100$ mK since
the electrons thermally decouple from the phonons leading to a saturation of the electron temperature \cite{Giazotto2006}.
%$V_{\rm B}$ can increase $T_{\rm N}$ because of the heat generated by the tunneling current.
Neverthless, the theoretical model can be used to convert the measured $V_{\rm th}$ to $T_{\rm N}$ even in this temperature range and under finite $V_{\rm B}$.
Figure~\ref{VT2}b shows the measured $T_{\rm N}$ for several bias voltages obtained using the calibration data in Figure~\ref{VT2}a.
%For the numerical simulation we linearly interpolate the data in Fig.~\ref{VT2}(b).

%%%%%%%%%%%%%%%%%%%%%%%%%%%%%%%%%%%%%%%%%%%%%%%%%%%%%%%%%%%%%%%%%%%%%
%% The "Acknowledgement" section can be given in all manuscript
%% classes.  This should be given within the "acknowledgement"
%% environment, which will make the correct section or running title.
%%%%%%%%%%%%%%%%%%%%%%%%%%%%%%%%%%%%%%%%%%%%%%%%%%%%%%%%%%%%%%%%%%%%%

%%%%%%%%%%%%%%%%%%%%%%%%%%%%%%%%%%%%%%%%%%%%%%%%%%%%%%%%%%%%%%%%%%%%%
%% The appropriate \bibliography command should be placed here.
%% Notice that the class file automatically sets \bibliographystyle
%% and also names the section correctly.
%%%%%%%%%%%%%%%%%%%%%%%%%%%%%%%%%%%%%%%%%%%%%%%%%%%%%%%%%%%%%%%%%%%%%
\bibliography{achemso-demo}

\begin{thebibliography}{99}
%\bibitem{Stockklauser2015} Stockklauser, A.; Maisi, V. F.; Basset, J.; Cujia, K.; Reichl, C.; Wegscheider, W.; Ihn, T.; Wallraff, A.; Ensslin, K., Phys. Rev. Lett. {\bf 2015} 115, 046802--046806.
%\bibitem{Koch2007} Koch, J.; Yu, T. M.; Gambetta, J.; Houck, A. A.; Schuster, D. I.; Majer, J.; Blais, A.; Devoret, M. H.;  Girvin, S. M.; Schoelkopf, R. J.; Phys. Rev. A {\bf 2007} 76, 042319--042337.
%\bibitem{Manucharyan2009} Manucharyan, V. E.;  Koch, J.;  Glazman, L. I.; Devoret, M. H.,  Science {\bf 2009} 326, 113--116.
%\bibitem{Lowell2013} Lowell, P. J.; Neil, G. C.; Underwood, J. M.; Ullom, J. N., Appl. Phys. Lett. {\bf 2013} 102, 082601--082604.
%\bibitem{Nataf2011} Nataf, P.; Ciuti, C., Phys. Rev. Lett. {\bf 2011}  107, 190402--190406.
\bibitem{Wolf2009} Wolf, E. L. {\it Quantum Nanoelectronics: An Introduction to Electronic
Nanotechnology and Quantum Computing} (Wiley-VCH, Weinheim, 2009).
\bibitem{Blais2004}  Blais, A., Huang, R. -S., Wallraff, A., Girvin, S. M. \& Schoelkopf, R. J. Cavity quantum electrodynamics for superconducting electrical circuits: An architecture for quantum computation. {\it Phys. Rev. A}  {\bf 69}, 062320 (2004).
\bibitem{Wallra2004} Wallraff, A., Schuster, D. I., Blais,  A., Frunzio, L.,  Huang, R.-S., Majer, J., Kumar, S., Girvin, S. M. \&  Schoelkopf, R. J. Strong coupling of a single photon to a superconducting qubit using circuit quantum electrodynamics. {\it Nature} {\bf 431}, 162--167 (2004).
\bibitem{Majer2007} 
Majer, J.,  Chow, J. M., Gambetta, J. M.,  Koch, J., Johnson, B. R.,  Schreier, J. A.,  Frunzio, L.,  Schuster, D. I., 
Houck, A. A.,  Wallraff, A., Blais, A., Devoret, M. H., Girvin, S. M. \& Schoelkopf, R. J. Coupling superconducting qubits via a cavity bus. {\it Nature} {\bf 449}, 443--447 (2007).
\bibitem{Sillanpaa2007} Sillanp\"a\"a, M. A., Park, J. I. \&  Simmonds, R. W. Coherent quantum state storage and transfer between two phase qubits via a resonant cavity. {\it Nature} {\bf 449}, 438--442 (2007).
\bibitem{Devoret2013}  Devoret, M. H. \&  Schoelkopf, R. J. Superconducting circuits for quantum information: an outlook. {\it Science} {\bf 339}, 1169--1174 (2013).
\bibitem{Kelly2015} Kelly, J. et al. State preservation by repetitive error detection in a superconducting quantum circuit. {\it Nature} {\bf 519}, 66--69 (2015).
\bibitem{Ofek2016} Ofek, N. et al. Extending the lifetime of a quantum bit with error correction in superconducting circuits. {\it Nature} {\bf 536}, 441--445 (2016).
\bibitem{Inomata2016} Inomata, K., Lin, Z. R., Koshino, K., Oliver, W. D., Tsai, J. S., Yamamoto, T. \& Nakamura, Y. Single microwave-photon detector using an artificial $\Lambda$-type three-level system. {\it Nat. Commun.} {\bf 7}, 12303 (2016).
\bibitem{Govenius2016} Govenius, J., Lake, R. E., Tan, K. Y. \& Möttönen, M. Detection of Zeptojoule microwave pulses using electrothermal feedback in proximity-induced Josephson junctions. {\it Phys. Rev. Lett.} {\bf 117}, 030802 (2016).
\bibitem{Saira2016} Saira, O.-P., Zgirski, M., Viisanen, K. L., Golubev, D. S. \& Pekola, J. P. Dispersive thermometry with a Josephson junction coupled to a resonator. {\it Phys. Rev. Applied} {\bf 6}, 024005 (2016).
\bibitem{Clark2005} Clark, A. M., Miller, N. A., Williams, S., Ruggiero, S. T., Hilton, G. C., Vale, L. R., Beall, K. D.,
Irwin, K. D. \& Ullom, J. N. Cooling of bulk material by electron-tunneling refrigerators. {\it Appl. Phys. Lett.} {\bf 86}, 173508--173510 (2005).
\bibitem{Timofeev2009} Timofeev, A. V.,  Helle, M., Meschke, M., M\"{o}tt\"{o}nen, M.\& Pekola, J. P.
Electronic refrigeration at the quantum limit. {\it Phys. Rev. Lett.} {\bf 102}, 200801 (2009).
\bibitem{Kuan2016} Tan, K. Y., Partanen, M., Lake, R. E., Govenius, J.,  Masuda, S. \& M\"{o}tt\"{o}nen, M.
Quantum-circuit refrigerator.  {\it arXiv}:1606.04728 (2016). 
%arXiv.org e-Print archive. https://arxiv.org/abs/1606.04728
%(accessed Jun 15, 2016).
\bibitem{Giazotto2006} Giazotto, F., Heikkil\"{a}, T. T., Luukanen, A., Savin, A. M. \& Pekola, J. P. Opportunities for mesoscopics in thermometry and refrigeration: Physics and applications. {\it Rev. Mod. Phys.}  {\bf 78}, 217--274 (2006).
\bibitem{Hofheinz2009} Hofheinz, M., Wang, H.,  Ansmann, M., Bialczak, R. C., Lucero, E., Neeley, M., O'Connell, A. D., Sank, D., Wenner, J., Martinis, J. M. \& Cleland, A. N. Synthesizing arbitrary quantum states in a superconducting resonator. {\it Nature} {\bf 459}, 546--549 (2009).
\bibitem{Meschke2006} Meschke, M., Guichard, W. \& Pekola, J. P. Single-mode heat conduction by photons. {\it Nature} {\bf 444}, 187--190 (2006).
\bibitem{Matti2016} Partanen, M., Tan, K. Y., Govenius, J., Lake, R. E., M\"{a}kel\"{a}, K., Tanttu, T. \& M\"{o}tt\"{o}nen, M. Quantum-limited heat conduction over macroscopic distances. {\it Nat. Phys.} {\bf 12}, 460--464 (2016).
\bibitem{Houck2007} Houck, A. A., Schuster, D. I., Gambetta, J. M., Schreier, J. A., Johnson, B. R., Chow, J. M., Frunzio, L., Majer, J., Devoret, M. H., Girvin, S. M. \& Schoelkopf, R. J. Generating single microwave photons in a circuit. {\it Nature} {\bf 449}, 328--331 (2007).
\bibitem{Peng2016} Peng, Z. H., de Graaf, S. E., Tsai, J. S. \& Astafiev, O. V. Tuneable on-demand single-photon source in the microwave range. {\it Nat. Commun.} {\bf 449}, 12588 (2016).
\bibitem{Pekola2010} Pekola, J. P., Maisi, V. F.,  Kafanov, S.,  Chekurov, N.,  Kemppinen, A.,
Pashkin, Y. A.,  Saira, O. -P.,  M\"{o}tt\"{o}nen M. \& Tsai, J. S. Environment-assisted tunneling as an origin of the Dynes density of states. {\it Phys. Rev. Lett.} {\bf 105}, 026803 (2010).
\bibitem{Devoret1990} Devoret, M. H., Esteve, D., Grabert, H., Ingold, G.-L.m Pothier, H. \& Urbina, C. Effect of the electromagnetic environment on the Coulomb blockade in ultrasmall tunnel junctions. {\it Phys. Rev. Lett.}  {\bf 64}, 1824 (1990).
\bibitem{Girvin1990} Girvin, S. M., Glazman, L. I., Jonson, M., Penn, D. R. \& Stiles, M. D. Quantum fluctuations and the single-junction Coulomb blockade. {\it Phys. Rev. Lett.} {\bf 64}, 3183 (1990).
\bibitem{Averin1990} Averin, B., Nazarov, Y. \& Odintsov, A. Incoherent tunneling of the cooper pairs and magnetic flux quanta in ultrasmall Josephson junctions. {\it Physica B} {\bf 165/166}, 945--946 (1990).
\bibitem{Ingold} Ingold, G. \& Nazarov, Y. V. Charge tunneling rates in ultrasmall junctions. {\it NATO ASI Series B} {\bf 294}, 21--107 (1992).
\bibitem{Bajjani2010} Zakka-Bajjani, E., Dufouleur, J., Coulombel, N., Roche, P., Glattli, D. C. \& Portier, F. Experimental determination of the statistics of photons emitted by a tunnel junction. {\it Phys. Rev. Lett.} {\bf 104}, 206802 (2010).
\bibitem{Hofheinz2011} Hofheinz, M., Portier, F., Baudouin, Q., Joyez, P., Vion, D., Bertet, P., Roche, P. \& Esteve, D. Bright side of the coulomb blockade. {\it Phys. Rev. Lett.} {\bf 106}, 217005 (2011).
\bibitem{You2007}  You, J. Q., Liu, Y. X., Sun, C. P. \& Nori, F. Persistent single-photon production by tunable on-chip micromaser with a superconducting quantum circuit. {\it Phys. Rev. B} {\bf 75}, 104516 (2007).
\bibitem{Astafiev2007} Astafiev, O., Inomata, K., Niskanen, A. O., Yamamoto, T., Pashkin, Y. A., Nakamura, Y. \& Tsai, J. S.  Single artificial-atom lasing. {\it Nature} {\bf 449}, 588--590 (2007).
\bibitem{Hauss2008}  Hauss, J., Fedorov, A., Hutter, C., Shnirman, A. \& Sch\"on, G. Single-qubit lasing and cooling at the Rabi frequency. {\it Phys. Rev. Lett.} {\bf 100}, 037003 (2008).
\bibitem{Grajcar2008}  Grajcar, M., van der Ploeg, S. H. W., Izmalkov, A., Ilichev, H. G. M. E., Fedorov, A., Shnirman, A. \& Sch\"on, G. Sisyphus cooling and amplification by a superconducting qubit. {\it Nat. Phys.} {\bf 4}, 612--616 (2008).
\bibitem{Bruhat2016} Bruhat, L. E., Viennot, J. J., Dartiailh, M. C., Desjardins, M. M., Kontos, T. \& Cottet, A. Cavity photons as a probe for charge relaxation resistance and photon emission in a quantum dot coupled to normal and superconducting continua. {\it Phys. Rev. X}  {\bf 6}, 021014 (2016).
\bibitem{Stockklauser2015} Stockklauser, A., Maisi, V. F., Basset, J., Cujia, K., Reichl, C., Wegscheider, W., Ihn, T., Wallraff, A. \& Ensslin, K. Microwave emission from hybridized states in a semiconductor charge qubit. {\it Phys. Rev. Lett.} {\bf 115}, 046802 (2015).
\bibitem{Childress2004} Childress, L.,  S\o{}rensen, A. S. \&  Lukin, M. D. Mesoscopic cavity quantum electrodynamics with quantum dots. {\it Phys. Rev. A} {\bf 69},
042302 (2004).
\bibitem{Liu2014} Liu, Y.-Y., Petersson, K. D., Stehlik, J., Taylor, J. M. \& Petta, J. R. Photon emission from a cavity-coupled double quantum dot. {\it Phys. Rev. Lett.} {\bf 113}, 036801 (2014).
\bibitem{Leivo1996} Leivo, M. M. \& Pekola, J. P. {\it Appl. Phys. Lett.} {\bf 68}, 1996--1998 (1996).
\bibitem{Gevorgian1995} Gevorgian, S., Linn\'{e}r, L. J. P. \&  Kollberg, E. L. CAD models for shielded multilayered CPW. {\it IEEE Trans. Microw. Theory Techn.} {\bf 43}, 772--779 (1995).
\bibitem{Goppl2008} G\"{o}ppl, M., Fragner, A.,  Baur, M., Bianchetti, R., Filipp, S., Fink, J. M.,  Leek, P. J., Puebla, G., Steffen, L. \& Wallraff. A. Coplanar waveguide resonators for circuit quantum electrodynamics. {\it J. Appl. Phys.} {\bf 104}, 113904--113911 (2008).










%\bibitem{Yin2013} Yin, Y.; Chen, Y.; Sank, D.;  O'Malley, P. J. J.; White, T. C.; Barends, R.;  Kelly, J. Lucero, E.; Mariantoni, M.; Megrant, A.; Neill, C.; Vainsencher, A.; Wenner, J.; Korotkov, A. N.; Cleland, A. N.; Martinis, J. M., Phys. Rev. Lett. {\bf 2013}, {110}, 107001--107005.
%\bibitem{Inomata2014} Inomata, K.; Koshino, K.; Lin, Z. R.; Oliver, W. D.; Tsai, J. S.; Nakamura, Y.; Yamamoto, T.,  Phys. Rev. Lett. {\bf 2014}, {113}, 063604--063608.
%\bibitem{Pechal2014} Pechal, M.; Huthmacher, L.;  Eichler, C.; Zeytino\u{g}lu, S.;  Abdumalikov, A. A.; Berger, Jr., S.; Wallraff, A.; Filipp, S.; Phys. Rev. X {\bf 2014} {4}, 041010--041018.


%\bibitem{Jones2013a} Jones, P. J.; Huhtam\"{a}ki, J. A.; M\"{o}tt\"{o}nen, M.; Sci. Rep. {\bf 2013}, {3}, 1987--1995.
%%

%%
%\bibitem{Frey2011}  T. Frey, P. J. Leek, M. Beck, K. Ensslin, A.Wallraff, and T. Ihn,
%Appl. Phys. Lett. {\bf 2011} 98, 262105--262107.
%We present measurements of a hybrid system consisting of a microwave transmission-line resonator and a lateral quantum dot defined on a GaAs heterostructure. The two subsystems are separately characterized and their interaction is studied by monitoring the electrical conductance through the quantum dot. The presence of a strong microwave field in the resonator is found to reduce the resonant conductance through the quantum dot and is attributed to electron heating and modulation of the dot potential.
%\textcolor{red}{maybe we can omit this.}
%%
%\bibitem{Delbecq2011} M. R. Delbecq, V. Schmitt, F. D. Parmentier, N. Roch, J. J. Viennot, G. F`eve, B. Huard, C. Mora, A. Cottet, and T. Kontos, Phys. Rev. Lett. {\bf 2011} 107, 256804--256808.
%We demonstrate a hybrid architecture consisting of a quantum dot circuit coupled to a single mode of the electromagnetic field. We use single wall carbon nanotube based circuits inserted in superconducting microwave cavities. By probing the nanotube dot using a dispersive readout in the Coulomb blockade and the Kondo regime, we determine an electron-photon coupling strength which should enable circuit QED experiments with more complex quantum dot circuits.
%% This is not photon source

%\bibitem{Frey2012} T. Frey, P. J. Leek, M. Beck, A. Blais, T. Ihn, K. Ensslin, and
%A. Wallraff, Phys. Rev. Lett. {\bf 2012} 108, 046807--046811.
%We demonstrate the realization of a hybrid solid-state quantum device, in which a semiconductor double quantum dot is dipole coupled to the microwave field of a superconducting coplanar waveguide resonator. The double dot charge stability diagram extracted from measurements of the amplitude and phase of a microwave tone transmitted through the resonator is in good agreement with that obtained from transport measurements.
%\bibitem{Jin2011} P.-Q. Jin, M. Marthaler, J. H. Cole, A. Shnirman, and G. Sch\"on,
%Phys. Rev. B {\bf 2011} 84, 035322--035330.
%We study a double quantum-dot system coherently coupled to an electromagnetic resonator. A current through the dot system can create a population inversion in the dot levels and, within a narrow resonance window, a lasing state in the resonator. The lasing state correlates with the transport properties. On one hand, this allows probing the lasing state via a current measurement.
%\bibitem{Jin2013}  J. Jin, M. Marthaler, P.-Q. Jin, D. Golubev, and G. Sch\"on, New
%J. Phys. {\bf 2013} 15, 025044--025058.
%Here we study the noise properties of the transport current in the lasing regime, i.e. both the zero-frequency shot noise and the noise spectrum.
%\bibitem{Bergenfeldt2012}  C. Bergenfeldt and P. Samuelsson, Phys. Rev. B {\bf 2012} 85, 045446--045462.
%We investigate theoretically the properties of the photon state and the electronic transport in a system consisting of a metallic quantum dot strongly coupled to a superconducting microwave transmission line cavity. 
%\bibitem{Schiro2014} M. Schir\'o and Karyn Le Hur, Phys. Rev. B {\bf 2014} 89, 195127--195137.
%These systems realize novel platforms to explore nonequilibrium quantum impurity physics with light and matter. Coupling the quantum dot system to reservoir leads (source and drain) produces an electronic current as well as dissipation when applying a bias voltage across the system. 
%\bibitem{Souquet2013} J. -R. Souquet, I. Safi, and P. Simon, Phys. Rev. B {\bf 2013} 88, 205419--205430.
%We study the out-of-equilibrium transport in a Tomonaga-Luttinger liquid containing a weak or a tunneling barrier coupled to an arbitrary electromagnetic environment. 


%\bibitem{Courtois2014} Courtois, H.; Hekking, F.; Nguyen, H.; Winkelmann, C., J. Low Temp. Phys. {\bf 2014} 175, 799--812.

%\bibitem{Sothmann2015} Sothmann, B.; S\'{a}nchez, R.; Jordan, A. N., Nanotechnology {\bf 2015}, {26}, 032001--032024.

%\bibitem{Srinivasan2011} Srinivasan, S. J.;  Hoffman, A. J.; Gambetta, J. M.; Houck, A. A.;
%Phys. Rev. Lett. {\bf 2011}, {106}, 083601--083604.
%\bibitem{Whittaker2014} J. D. Whittaker et al., Phys. Rev. B {\bf 2014}, 90, 024513--024527.

%\bibitem{Jones2013} P. J. Jones, J. Salmilehto and M. M\"{o}tt\"{o}nen, J. Low. Temp. Phys. {\bf 173}, 17 (2013).
%\bibitem{Clerk2010} A. A. Clerk, M. H. Devoret, S. M. Girvin, F. Marquardt and R. J. Schoelkopf, Rev. Mod. Phys. {\bf 82}, 1155 (2010).
%\bibitem{Nakamura1999} Nakamura, Y., Nature {\bf 1999}, {398}, 786--788.
%\bibitem{Brouri2000} Brouri, R.; Beveratos, A.; Poizat, J. P.; Grangier, P.;  Opt. Lett. {\bf 2000}, {25}, 1294--1296.
%\bibitem{Kurtsiefer2000} Kurtsiefer, C.; Mayer, S.; Zarda, P.; Weinfurter, H., Phys. Rev. Lett. {\bf 2000}, {85}, 290--293.
%\bibitem{Rabeau2005} Rabeau, J. R.; Chin, Y. L.; Prawer, S.; Jelezko, F.; Gaebel, T.; Wrachtrup, J.; Appl. Phys. Lett. {\bf 2005}, {86}, 131926--131928.
%\bibitem{Gerard1999} G{\'e}rard, J. M.; Gayral, B., J., Lightwave Technol. {\bf 1999}, {17}, 2089--2095.
%\bibitem{Moreau2001} Moreau, E.; Robert, I.; G{\'e}rard, J. M.; Abram, I.; Manin, L.; Thierry-Mieg, V., Appl. Phys. Lett., {\bf 2001}, {79}, 2865--2868.
%\bibitem{Michler2000} Michler, P.; Imamoglu, A.; Mason, M. D.; Carson, P. J.; Strouse, G. F.; Buratto, S. K., Nature {\bf 2000}, {406}, 968--970.
%\bibitem{Santori2001} Santori, C.; Pelton, M.; Solomon, G.; Dale, Y.; Yamamoto, Y., Phys. Rev. Lett. {\bf 2001}, {86}, 1502--1505.
%\bibitem{Claudon2010} Claudon, J.; Bleuse, J.; Malik, N. S.; Bazin, M.; Jaffrennou, P.; Gregersen, N.; Sauvan, C.; Lalanne, P.; Gerard, J. -M., Nature Photonics {\bf 2010}, {4}, 174--177.
%\bibitem{Heindel2010} Heindel, T.; Schneider, C.; Lermer, M.; Kwon, S. H.; Braun, T.; Reitzenstein, S.; Hö{\"e}fling, S.;  Kamp, M.; Forchel, A., Appl. Phys. Lett., {\bf 2010},  {96} 011107--011109. 

%\bibitem{Nowak2014} Nowak, A. K.;  Portalupi, S. L.;  Giesz, V.;  Gazzano, O.; 	Dal Savio, C.;  Braun, P.-F.; Karrai, K.;  Arnold, C.; Lanco, L.; Sagnes, I.; Lema\^{i}tre, A.; Senellart, P., Nat. Commun. {\bf 2014}  {5} 439--446.

%\bibitem{Yuan2002} 
%Yuan, Z.; Kardynal, B. E.; Stevenson, R. M.;
%Shields, A. J.; Lobo, C. J.; Cooper, K.;
%Beattie, N. S.; Ritchie, D. A.; Pepper, M., Science, {\bf 2002}, {295}, 102--105.
%bibitem{Kuhn2002} Kuhn, A.; Hennrich, M.; Rempe, G., Phys. Rev. Lett. {\bf 2002}, {89}, 067901--067904.
%\bibitem{Pascal2011} Pascal, L. M. A.; Courtois, H.; Hekking, F. W. J.; Phys. Rev. B {\bf 2011}, {83}, 125113--125119.
%\bibitem{Jones2012A} Jones, P. J.; Huhtam\"{a}ki, J. A. M.; Partanen, M.; Tan, K. Y.; M\"{o}tt\"{o}nen, M.;  Phys. Rev. B {\bf 2012}, {86}, 035313--035320.
%\bibitem{Pekola2015} Pekola, J. P.,  Nature Phys. {\bf 2015}, 11, 118--123 (2015).
%\bibitem{Golubev2015} Golubev, D. S.; Pekola, J. P.; Phys. Rev. B {\bf 2015}, {92}, 085412--085418.
%\bibitem{Hofheinz2009} Hofheinz, M.; Wang, H.;  Ansmann, M.; Bialczak, R. C.; Lucero, E.; Neeley, M.; O'Connell, A. D.; Sank, D.; Wenner, J.; Martinis, J. M.; Cleland, A. N., Nature {\bf 2009}, {459}, 546--549.

%\bibitem{Srinivasan2014}  Srinivasan, S. J.;  Sundaresan, N. M.; Sadri, D.; Liu, Y.; Gambetta, J. M.; Yu,  T.,
%Phys. Rev. A {\bf 2014}, {89}, 033857--033861.
%\bibitem{Pierre2014} Pierre, M.; Svensson, I.-M.; Sathyamoorthy, S. R.;  Johansson, G.; Delsing, P.,
%Appl. Phys. Lett. {\bf 2014}, {104}, 232604--232607.


\end{thebibliography}

\clearpage

\clearpage

%\clearpage
%\begin{figure}
%\begin{center}
%\includegraphics[width=9cm]{TOC_10_4_16.eps}
%\end{center}
%\caption*{Graphical TOC Entry.}
%\end{figure}

\end{document}